\documentclass[aps,prb,reprint,superscriptaddress]{revtex4-2}

\usepackage[english]{babel}
\usepackage[utf8x]{inputenc}
\usepackage[T1]{fontenc}
\usepackage{siunitx}
\usepackage{xcolor}
\usepackage[version=4]{mhchem}

\usepackage{longtable}
\usepackage{booktabs}
\usepackage{bibentry}
\usepackage{amsmath}
\usepackage{graphicx}
\usepackage[nolist]{acronym}

\usepackage[colorlinks=true, allcolors=blue,breaklinks=true]{hyperref}

\usepackage{ragged2e}
\usepackage{makecell}

\DeclareSIUnit\plus{+}

\newcommand{\beginsupplement}{%
        \setcounter{table}{0}
        \renewcommand{\thetable}{S\Roman{table}}%
        \setcounter{figure}{0}
        \renewcommand{\thefigure}{S\arabic{figure}}%
        \setcounter{subsection}{0}
        \renewcommand{\thesubsection}{S-\Alph{subsection}}%
        \setcounter{subsubsection}{0}
        \renewcommand{\thesubsubsection}{\arabic{subsubsection}}%
     }
     
\newcommand{\beginappendix}{%
        \setcounter{table}{0}
        \renewcommand{\thetable}{A\Roman{table}}%
        \setcounter{figure}{0}
        \renewcommand{\thefigure}{A\arabic{figure}}%
        \setcounter{subsection}{0}
        \renewcommand{\thesubsection}{Appendix \Alph{subsection}}%
     }

\begin{document}

\title{Sub-megahertz homogeneous linewidth for Er in Si via in situ single photon detection}
\author{Ian R. \surname{Berkman}}
\email{i.berkman@unsw.edu.au}
\affiliation{Centre of Excellence for Quantum Computation and Communication Technology, School of Physics, University of New South Wales, Sydney, NSW 2052, Australia}
\author{Alexey \surname{Lyasota}}
\affiliation{Centre of Excellence for Quantum Computation and Communication Technology, School of Physics, University of New South Wales, Sydney, NSW 2052, Australia}
\author{Gabriele G. \surname{de Boo}}
\affiliation{Centre of Excellence for Quantum Computation and Communication Technology, School of Physics, University of New South Wales, Sydney, NSW 2052, Australia}
\author{John G. \surname{Bartholomew}}
\affiliation{Centre for Engineered Quantum Systems, School of Physics, The University of Sydney, Sydney, NSW 2006, Australia}
\affiliation{The University of Sydney Nano Institute, The University of Sydney, Sydney, NSW 2006, Australia}
\author{Brett C. \surname{Johnson}}
\affiliation{Centre of Excellence for Quantum Computation and Communication Technology, School of Physics, University of Melbourne, Victoria 3010, Australia}
\affiliation{Centre of Excellence for Quantum Computation and Communication Technology, School of Engineering, RMIT University, Victoria 3001, Australia}
\author{Jeffrey C. \surname{McCallum}}
\affiliation{Centre of Excellence for Quantum Computation and Communication Technology, School of Physics, University of Melbourne, Victoria 3010, Australia}
\author{Bin-Bin \surname{Xu}}
\affiliation{Centre of Excellence for Quantum Computation and Communication Technology, School of Physics, University of New South Wales, Sydney, NSW 2052, Australia}
\author{Shouyi \surname{Xie}}
\affiliation{Centre of Excellence for Quantum Computation and Communication Technology, School of Physics, University of New South Wales, Sydney, NSW 2052, Australia}
\author{Rose L. \surname{Ahlefeldt}}
\affiliation{Centre of Excellence for Quantum Computation and Communication Technology, Research School of Physics, Australian National University, Canberra, ACT 0200, Australia}
\author{Matthew J. \surname{Sellars}}
\affiliation{Centre of Excellence for Quantum Computation and Communication Technology, Research School of Physics, Australian National University, Canberra, ACT 0200, Australia}
\author{Chunming \surname{Yin}} %\orcid{0000-0003-0117-8225}
\affiliation{Centre of Excellence for Quantum Computation and Communication Technology, School of Physics, University of New South Wales, Sydney, NSW 2052, Australia}
\affiliation{%
Hefei National Laboratory for Physical Sciences at the Microscale, CAS Key Laboratory of Microscale Magnetic Resonance and School of Physical Sciences, University of Science and Technology of China, Hefei 230026, China
}
\author{Sven \surname{Rogge}}
\affiliation{Centre of Excellence for Quantum Computation and Communication Technology, School of Physics, University of New South Wales, Sydney, NSW 2052, Australia}

% Situation
% =========
% \ce{Er} in \ce{Si} is a good candidate for quantum information purposes
% \ce{Er} has a good wavelength, long coherence and microwave transitions
% \ce{Si} is fab friendly and provides a low magnetic noise environment

% Complication
% ============
% We haven't identified \ce{Er} sites in \ce{Si} with the right properties (inhomogeneous broadening, homogeneous broadening / spectral diffusion + coherence, spin transition addressable, nuclear spin accessible, radiative efficiency)
% Interaction of optical emitters with local and free charges in semiconductors leads to
%		- nonradiative recombination -> reduction in luminescence, lifetime
%		- spectral diffusion (See Anderson 2019)

% Question
% ========
% Can we identify suitable sites in \ce{Si} for QI?

% Answer
% ======
% We find many optical transitions with promising properties
%   linewidth, lifetime, spectral diffusion, Zeeman

% Discuss \ce{Er} in YSO results that demonstrate the potential of \ce{Er} for quantum information

%\ce{Er} in orthosilicate (YSO) ...
%homogeneous linewidth of \SI{73}{\hertz} \cite{bottger2009}
%single ions \cite{dibos2018}
%\cite{chen2020}
%spin state read-out of single ions \cite{raha20}
%Coupling to superconducting resonators? \cite{bushev2011}

%%%%%%%%%%%%%%%%%%%%%%%%%%%%%%%%%%%%%%%%%%%%%%%%%%%%%%%%%%%%%%%%%%%%%%%%%%%%%%%

\begin{abstract}

We studied the optical properties of a resonantly excited trivalent \ce{Er} ensemble in \ce{Si} accessed via in situ single photon detection.
A novel approach which avoids nanofabrication on the sample is introduced, resulting in a highly efficient detection of \num{70} excitation frequencies, of which \num{63} resonances have not been observed in literature. 
The center frequencies and optical lifetimes of all resonances have been extracted, showing that \SI{5}{\percent} of the resonances are within \SI{1}{\giga\hertz} of our electrically detected resonances and that the optical lifetimes range from \SI{0.5}{\milli\second} up to \SI{1.5}{\milli\second}.
We observed inhomogeneous broadening of less than \SI{400}{\mega\hertz} and an upper bound on the homogeneous linewidth of \SI{1.4}{\mega\hertz} and \SI{0.75}{\mega\hertz} for two separate resonances, which is a reduction of more than an order of magnitude observed to date.
These narrow optical transition properties show that \ce{Er} in \ce{Si} is an excellent candidate for future quantum information and communication applications.

\end{abstract}

\maketitle

\begin{acronym}
    \acro{RE}[RE]{rare earth}
	\acro{PL}[PL]{photoluminescence}
	\acro{PLE}[PLE]{photoluminescence excitation}
	\acro{EPR}[EPR]{electron paramagnetic resonance}
%	\acro{EL}[EL]{electroluminescence}
%	\acro{CR}[CR]{count rate}
%	\acro{DCR}[DCR]{dark count rate}
	\acro{SSPD}[SSPD]{superconducting single photon detector}
	\acro{SNR}[SNR]{signal-to-noise ratio}
	\acro{AOM}[AOM]{acousto-optical modulator}
	\acro{EOM}[EOM]{electro-optical modulator}
%	\acro{HEMT}[HEMT]{high-electron-mobility transistor}
%	\acro{ZEFOZ}[ZEFOZ]{zero first-order Zeeman}
%	\acro{CMOS}[CMOS]{complementary metal-oxide-semiconductor}
	\acro{FinFET}[FinFET]{fin field-effect transistor}
%	\acro{PECVD}[PECVD]{plasma-enhanced chemical vapour deposition}
	\acro{FWHM}[FWHM]{full width at half maximum}
\end{acronym}

%%%%%%%%%%%%%%%%%%%%%%%%%%%%%%%%%%%%%%%%%%%%%%%%%%%%%%%%%%%%%%%%%%%%%%%%%%%%%%%

\section*{Introduction} 
\Ac{RE} ions embedded in a host crystal possess numerous interesting properties for quantum information processing. \Ac{RE} ions can have near-lifetime limited coherence times on their optical transitions \cite{thiel2011}, as long as \SI{4.4}{\milli\second} \cite{Bottger2009}, and hyperfine coherence times from seconds to hours in carefully controlled magnetic fields \cite{Zhong2015, Rancic2017}. 
While \ac{RE} ions have weaker oscillator strengths than other solid state optical emitters, single ion readout has been achieved in a multiple of host crystals such as \ce{YAlO_3}\cite{kolesov2012,siyushev2014}, \ce{YVO_4}\cite{kindem2020}, \ce{Y_2SiO_5}\cite{raha20, chen2020, utikal2014} and \ce{Si}~\cite{yin13}. These properties make \acs{RE} ions in solid state hosts excellent material for quantum memories\cite{zhong2017, Zhong2015} and potential candidates for future qubits~\cite{grimm2021}.

The \ac{RE} ion \ce{Er} in solids most commonly has a \ce{^4I_{15/2}} ground state and a \ce{^4I_{13/2}} excited state~\cite{kenyon2005}. 
The optical transition from the lowest crystal field level of \ce{^4I_{15/2}} ground state to the lowest lying level of the \ce{^4I_{13/2}} excited state occurs at approximately \SI{1540}{\nano\metre}, hence within the technologically important telecom C-band. This convenient wavelength makes \ce{Er} particularly attractive for quantum communication applications, as \ce{Er}-based devices will be telecom-compatible.

Incorporating \ce{Er} into \ce{Si}, by means such as ion implantation or chemical vapour deposition, allows integration into standard complementary metal-oxide-semiconductor processing and provides the ability to fabricate nanophotonic structures~\cite{weiss21}. 
Furthermore, \ce{Si} can be enriched to less than \num{1} ppm \ce{^{29}Si}, effectively resulting in a low magnetic noise environment \cite{steger2012, muhonen2014} which leads to linewidths as narrow as \SI{33}{\mega\hertz} for T centers \cite{bergeron2020} and \SI{5}{\mega\hertz} for the donor-bound excitons~\cite{yang2009} in silicon. 

%%%%%%%%%%%%%%%%%%%%%%%%%%%%%%%%%%%%%%%%%%%%%%%%%%%%%%%%%%%%%%%%%%%%%%%%%%%%%%%

\section*{Resonant photoluminescence excitation spectroscopy}

\ce{Er} in \ce{Si} is able to occupy multiple classes of sites \cite{kenyon2005, przybylinska1996}, however most of the sites essential for quantum information have not been characterized thus far. The three main methods used to identify these \ce{Er}:\ce{Si} sites are namely \ac{EPR}, \ac{PL} and \ac{PLE}. 

Methods utilizing \Ac{EPR} rely on microwaves to probe the paramagnetic properties of electrons in closely separated levels within the $^4I_{15/2}$ state. 
However, the use of microwaves makes it impossible to excite from the \ce{^4I_{15/2}} ground state to the \ce{^4I_{13/2}} excited state. 
Both \ac{PL} and \ac{PLE} address these energy levels by using photon excitation. In a \ac{PL} experiment, above-bandgap light is used to excite the \ce{Er} ions and the photoluminescence is recorded using a spectrometer.
The spectrum includes decay from the lowest \ce{^4I_{13/2}} to the multiple \ce{^4I_{15/2}} levels following the excitation of the \ce{Si} host~\cite{przybylinska1996}, where the intensity depends on the excitation transfer efficiency from the \ce{Si}~\cite{hogg1996}. 
The crystal field levels in the excited \ce{^4I_{13/2}} state are inaccessible in \ac{PL}, but obtainable in \ac{PLE} when using a narrow band laser to resonantly excite the population into multiple \ce{^4I_{13/2}} levels and collect the photoluminescence. 
In addition, less free carriers are generated that can affect the spectrum and lifetime~\cite{Palm1996}. 
Differences of exciting resonantly or using above-band gap excitation have been observed in \ce{Er}-doped \ce{GaAs} where \ce{Er} centers that exhibited photoluminescence only under direct \ce{4f}-shell excitation did not show photoluminescence under above-band gap excitation~\cite{takahei1995}, thus corroborating the significance of \ac{PLE} for investigating photoluminescent \ce{Er} sites and the corresponding \ce{^4I_{13/2}} crystal field levels in semiconductors.

Recent \ac{PLE} experiments in \ce{Er}:\ce{Si} have shown nine narrow photoluminescence resonances associated with different \ce{Er} sites and potentially different crystal field levels of these sites~\cite{weiss20arXiv}. 
Our measurement of a higher density sample observed \num{7} of the previously detected resonances, marked by the $\dagger$ in Table \ref{tab:resonances}, and characterized \num{63} additional resonances.

%%%%%%%%%%%%%%%%%%%%%%%%%%%%%%%%%%%%%%%%%%%%%%%%%%%%%%%%%%%%%%%%%%%%%%%%%%%%%%%

\subsection*{Experiment}
In our experiment we collected the emission of \ce{Er^{3+}} ions in an \SI{1.7}{\milli\metre}$\times$\SI{1.7}{\milli\metre} \ce{Si} chip following resonant excitation with a laser. 
The chip was diced from a \SI{300}{\micro\metre} thick double-side-polished \ce{Si} wafer containing a background doping of \ce{P} ranging between \SIrange{0.9e15}{5e15}{\centi\metre\tothe{-3}}. 
To study the optical transitions without the complication of hyperfine splitting, the nuclear spin-free \ce{^{170}Er} isotope was implanted with multiple ion energies and fluences into one side of the chip to form a constant concentration profile of \SI{1e18}{\centi\metre\tothe{-3}} over a depth of \SI{0.2}{\micro\metre} to \SI{0.6}{\micro\metre}. 
The \ce{Er} concentration of the sample is an order of magnitude higher than the concentration used in our electrical detection experiments \cite{yin13,deboo20} to potentially increase the probability of detecting photoluminescence. 
In \ac{PL} experiments, co-implanting \ce{Er} with \ce{O} increases the photoluminescence \cite{michel1991,coffa1993} and leads to sharp lines in \ac{EPR} spectroscopy~\cite{carey1996}. 
Hence, \ce{O} was likewise implanted with multiple energies to create an overlapping profile but with a concentration of \SI{1e19}{\centi\metre\tothe{-3}}. 
Following implantation, the chip was annealed at \SI{700}{\celsius} for \SI{10}{\minute} in an \ce{N_2} atmosphere, which has resulted in optically active \ce{Er} ions in silicon \cite{yin13, weiss21}. 
Afterwards, a \SI{190}{\nano\metre} thick \ce{SiN_x} anti-reflective coating on both sides was formed using plasma-enhanced chemical vapor deposition to reduce Fabry-P\'{e}rot oscillations and enhance the optical transmission through the \ce{Si}. 

% FYI: Implant parameters (March 2018):
% Er(0.8 MeV, 0.9e13)+Er(1.2 MeV, 1.4e13)+Er(2.0 MeV, 3.1e13)
% O(100 keV, 1.1e14)+O(160 keV, 1.5e14)+O(250 keV, 2.4e14)

The implanted side was placed against the top of a \ce{WSi}-based, optical cavity-embedded \ac{SSPD} [Supplemental Material S-A]. 
\acp{SSPD} are able to detect single photons with high efficiency in a wide spectral range from \SI{250}{\nano\metre} \cite{wollman2017} up to \SI{7}{\micro\metre} \cite{goltsman2007,marsili2012,chen2017}. By adjusting the thickness of the optical cavity, the absorption can be tailored to a desired wavelength. 
We fabricated an \ac{SSPD} with a peak system detection efficiency of \SI{66.27}{\percent} at \SI{1550}{\nano\metre} including a bandwidth of approximately \SI{\pm50}{\nano\metre} for the measurement. 

The chip was sandwiched between the \ac{SSPD} and an optical fiber that had its core aligned with the \ac{SSPD}. 
The latter allowed an optical excitation of \ce{Er} ions in direct proximity to the \ac{SSPD}. 
The optical mode waist resulted in a tightly focused excitation spot at the \ce{Er} rich sample plane of \SI{20}{\micro\meter}, equal to the working area of the \ac{SSPD}. 
The experiment was operated at \SI{300}{\milli\kelvin} to ensure a low dark count rate on the \ac{SSPD}. 
Moreover, this temperature is low enough to minimize non-radiative recombination of the \ce{Er}~\cite{Palm1996,Taguchi1998,Priolo1998}.

To excite the \ce{Er} ions, we used a semiconductor diode laser (Pure Photonics PPCL550) with an output pulse modulated by two \acp{AOM} connected in series, resulting in an extinction ratio greater than \SI{100}{\decibel}.
After the excitation pulse, we recorded the number of counts from the \ac{SSPD} with a digital counter (Keysight 53131A or National Instruments PCI-6602).

%Additionally, the resolution of resonant \ac{PLE} is limited by the wavelength meter (\SI{\pm 0.8}{\pico\metre} for the Bristol 671B) whereas the resolution of the spectra obtained from above-band gap photoexcitation is limited by the resolution of the spectrometer.

%%%%%%%%%%%%%%%%%%%%%%%%%%%%%%%%%%%%%%%%%%%%%%%%%%%%%%%%%%%%%%%%%%%%%%%%%%%%%%%

\subsection*{Broad spectral survey} 

\begin{figure}
	\centering
    \includegraphics[width=1.0\columnwidth]{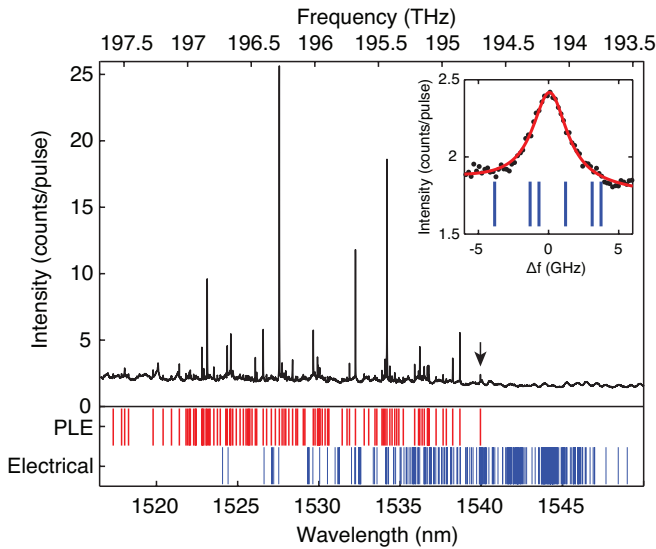}
    \caption{The \ac{PLE} spectrum compared with the single \ce{Er} ions found by electrical detection as function of the wavelength measured in vacuum using the wavemeter. 
    The red lines below the plot indicate the center of optically detected resonances and the blue lines indicate the center of electrically detected resonances. 
    Inset: a single optically detected inhomogeneous resonance centered at \SI{1539.948}{\nano\metre} which is fit with a Lorentzian line shape, resolving in a \ac{FWHM} of \SI{1.5}{\giga\hertz}. 
    The blue lines indicate \num{6} electrically detected resonances less than \SI{5}{\giga\hertz} away from the centre. 
    The \ac{FWHM} of the electrically detected resonances are less than \SI{100}{\mega\hertz}.}
    \label{fig:spectrum}
\end{figure}

The resonant \ac{PLE} spectrum was obtained by pulsing the laser for \SI{100}{\micro\second} and integrating the counts from \SI{10}{\micro\second} to \SI{1}{\milli\second} after the pulse [Fig. \ref{fig:spectrum}]. 
This was repeated \num{1000} times at each optical frequency before the excitation laser was stepped to the next optical frequency. 
In total, the range of \SIrange{1516}{1550}{\nano\metre} has been scanned in steps of \SI{50}{\mega\hertz} (\SI{0.4}{\pico\metre}). 
The laser line was broadened with a frequency modulation of \SI{60}{\mega\hertz} in order to avoid stepping over narrow resonances. 
The laser frequency was monitored at each step with a wavemeter (Bristol 621B).

%\section{Results}

%Mention that it is measured with the wavemeter

The spectrum in figure \ref{fig:spectrum} consists of \num{70} peaks which all displayed a prominence of at least \num{0.15} counts per pulse, followed a distribution profile and showed an exponential lifetime decay. 
The excitation wavelength, amplitude, Lorentzian \ac{FWHM} and optical lifetime of the resonances are listed in the appendix \ref{tab:resonances} and the raw data of the spectrum is provided in the Supplemental Material. 
These resonances are consistent with different \ce{Er} sites and may include resonances associated with excitation to higher crystal field levels of the \ce{^4I_{13/2}} manifold. 
Numerous sites could have been activated by the co-implanted \ce{O} due to the formation of \ce{Er}-\ce{O} complexes~\cite{kenyon2005}. 
The observed resonances had linewidths comparable to \ce{Er} in other host crystals\cite{dibos2018,bottger2006} and can be attributed to different environmental inhomogeneities as well as different sensitivities of the optical transition to these inhomogeneities.

In Fig. \ref{fig:spectrum}, the spectrum is compared to the resonantly optically excited, electrically detected resonances of single \ce{Er} ions \cite{yin13,deboo20}, indicated by the vertical blue lines, where those excitations were detected via ionization of a nearby charge trap in a \ce{Si} \ac{FinFET} device.  
Electrical detection resulted in numerous detected resonances at longer wavelengths than \SI{1540}{\nano\metre} which are not observed in the \ac{PLE} spectrum of the current sample. 
Inherently, optical and electrical detection rely on different decay mechanisms. Optical detection favors sites with a relatively high probability of radiative decay, whereas electrical detection relies on non-radiative decay processes of \ce{Er} sites. 
The electrically detected spectrum is the resulting histogram from multiple \ac{FinFET} devices with varying wavelength scans, different channel dimensions, background doping, \ce{Er} densities and at temperatures ranging from \SI{20}{\milli\kelvin} to \SI{4}{\kelvin}. 
In total, we found \SI{5}{\percent} of the optically detected \ac{PLE} resonances are within \SI{1}{\giga\hertz} of the electrically detected resonances, which indicate that they could originate from the same site. This could be confirmed by measuring g-tensors in both sets of sites.

\begin{figure}
	\centering
        \includegraphics[width=1.0\columnwidth]{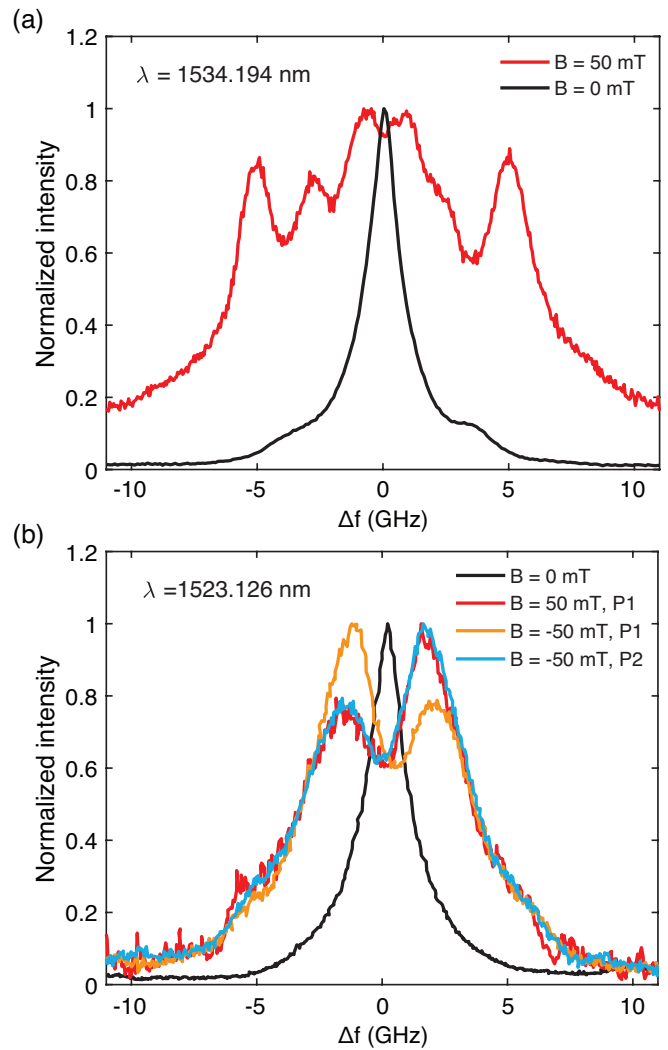}
        \caption{(a) Inhomogeneously broadened resonance at \SI{1534.194}{\nano\metre} at zero field and the splitting into multiple resonances when \SI{50}{\milli\tesla} is applied. 
        (b) Inhomogeneously broadened resonance at \SI{1523.126}{\nano\metre} under \SIlist{-50;0;50}{\milli\tesla}. 
        P1 indicates the polarization which resulted in a maximum intensity of the left peak at \SI{-50}{\milli\tesla} while P2 indicates the polarization which resulted in a maximum intensity of the right peak at \SI{-50}{\milli\tesla}.}
        \label{fig:magneticfield}
\end{figure}

%Mention orientation of chip in B field

In the next experiment, the splitting of inhomogeneously broadened resonances under an applied magnetic field was studied to understand their site symmetry.
The magnetic field was applied perpendicularly to the \ac{SSPD} and sample, and was limited to \SI{60}{\milli\tesla} before the \ac{SSPD} transitioned from a superconducting to normal state. This caused the \ac{SSPD} to be incapable of detecting single photons.
Under this magnetic field, the resonances presented in following section that showed a narrow spectral hole did not split sufficiently, indicating a small g-factor along the direction of the magnetic field.

Two resonances are presented in Fig. \ref{fig:magneticfield}(a) and Fig. \ref{fig:magneticfield}(b) which showed different site symmetries.
\ce{Er} ions in sites below cubic symmetry retain a two-fold Kramers' degeneracy in zero field, which is lifted in an applied magnetic field. 
This in total gives rise to four different possible optical transitions, assuming the g factors in the \ce{^4I_{15/2}} and \ce{^4I_{13/2}} differ. 
In the case where \ce{Er} is located in a site with a point symmetry lower than a cubic symmetry of the Si crystal, the number of magnetically inequivalent subsites can increase up to \num{24}, which corresponds to the C$_1$ point symmetry. 
Splitting of an inhomogeneously broadened resonance into a multiple of well distinguished lines thus confirms that the that the \ce{Er} centers reside in well defined crystallographic sites~\cite{vinh04}.

In Fig. \ref{fig:magneticfield}(a) a resonance is showed which splits in six lines, which is explained by magnetically inequivalent sites at a lower site symmetry than a cubic symmetry. In Fig. \ref{fig:magneticfield}(b) a resonance  at \SI{1523.199}{\nano\metre} is presented which split in two Zeeman arms with different intensities. 
The two Zeeman arms were broader than the resonance at zero field, indicating that the remaining two or more lines have not been split sufficiently. 
The split resonance showed asymmetric peak intensities, which can be attributed to different polarisation dependent oscillator strengths, rather than different Boltzmann populations of the initial states, because the relative intensities of the two peaks can be reversed by rotating the polarisation using a $\lambda/2$ waveplate. 
Reversing the magnetic field also reversed the peak intensities, further confirming that the difference is not due to differing Boltzmann populations.

%%%%%%%%%%%%%%%%%%%%%%%%%%%%%%%%%%%%%%%%%%%%%%%%%%%%%%%%%%%%%%%%%%%%%%%%%%%%%%%

%%%%%%%%%%%%%%%%%%%%%%%%%%%%%%%%%%%%%%%%%%%%%%%%%%%%%%%%%%%%%%%%%%%%%%%%%%%%%%%
%Mention instantaneous spectral diffusion
\subsection*{Homogeneous broadening}

\begin{figure}
	\centering
        \includegraphics[width=1.0\columnwidth]{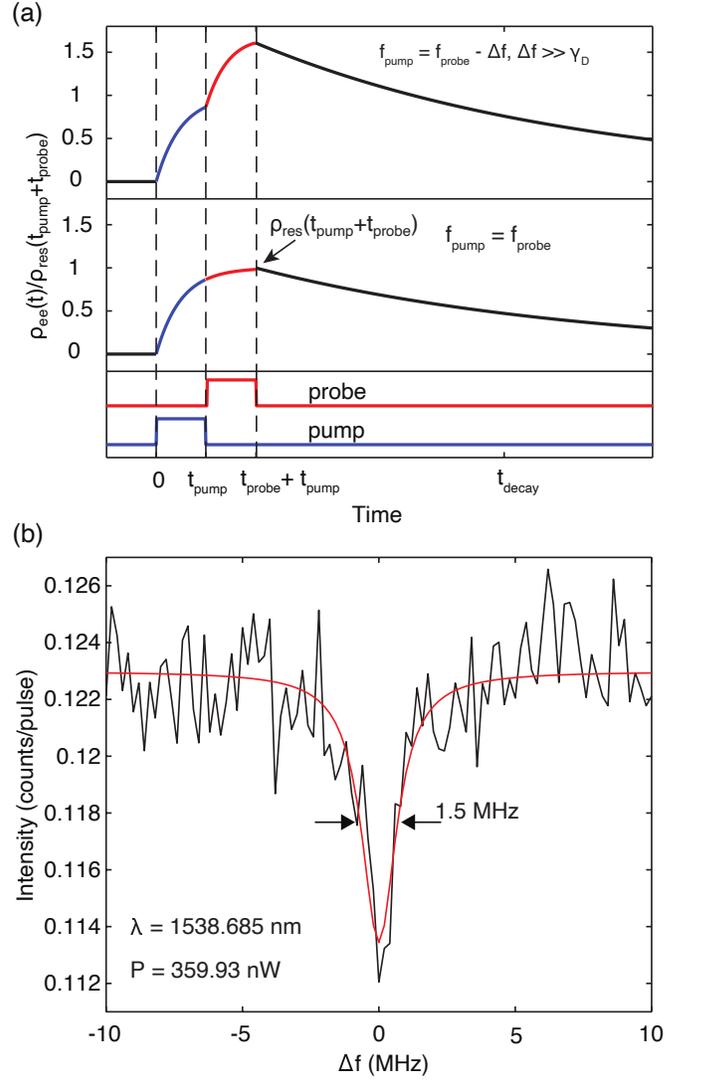}
        \caption{(a) The pulse schematic of the pump and probe microwave sources. The top curve represents the decay signal when the probe frequency is larger than the homogeneous line width. The lower curve represents the decay signal when the probe frequency is equal to the probe frequency, effectively resulting in a pulse of \num{2}$\text{t}_\text{pump}$ at $\text{f}_\text{pump}$ which leads to a reduced decay signal. The curves have been normalized to $\rho_{res}$ which is the occupation number when the subset is resonantly excited. (b) Upper bound on the homogeneous linewidth at \SI{1538.685}{\nano\metre}. The integrated number of counts up to \SI{1}{\milli\second} after the probe pulse as function of $\text{f}_\text{probe} - \text{f}_\text{pump}$. The data is fitted with a Lorentzian distribution, resulting in a FWHM of \SI{1.5}{\mega\hertz}.}
        \label{fig:HL}
\end{figure}

The homogeneous linewidth was investigated using transient spectral hole burning~\cite{Szabo1975}.
The method relied on the saturation of the fluorescence of an optical transition when the excitation pulse length exceeds the dephasing time due to the saturation of the atomic transition. 
The optical laser frequency was modulated using an \ac{EOM}, creating two sidebands \SI{5}{\giga\hertz} apart while suppressing the carrier. 
The high frequency sideband is centered on the inhomogeneous peak and excites the ensemble for \SI{20}{\micro\second} followed by a \SI{20}{\micro\second} excitation at a detuned laser frequency ($\Delta f$), referred to as the pump and probe pulse respectively.
For an equal pump and probe time, the occupation number of the excited state at the end of both pulses is given by

\begin{equation}
    \rho_{ee}(2t_\text{p}) = 
    \begin{cases}
        \rho_{res}(2t_p) & \text{for } \Delta f = 0\\    
        2\rho_{res}(t_p)(1-\frac{1}{2}e^{-t_p/\tau}) & \text{for } \Delta f \gg \gamma_D
    \end{cases},
\end{equation}
where $t_p$ is the pump time, $\rho_{res}$ is the occupation number when excited on resonance, $\tau$ the optical lifetime and $\gamma_D$ the homogeneous linewidth. 
The excitation and probe pulse length were chosen to be sufficiently short compared to the optical lifetime resulting in $\rho_{ee}(2t_\text{pump}) \approx 2\rho_\text{res}(t_\text{pump}) > \rho_{res}(2t_\text{pump})$ whenever $\Delta f \gg \gamma_D$. 
The repetition time of \SI{3}{\milli\second} was chosen to be twice the optical lifetime, ensuring the majority of excited \ce{Er} ions have decayed into the ground states. 

To ensure that the off-resonant low frequency sideband did not affect the spectral hole width, the homogenenous broadening was remeasured while the carrier was present. 
The results present a comparable width when the carrier is present, concluding that hole width is unaffected by the off-resonant light.

Under \SI{360}{\nano\watt} of excitation power, a spectral hole was visible at \SI{1538.685}{\nano\metre} and \SI{1532.254}{\nano\metre}. 
The data is fitted with a Lorentzian distribution resulting in a \ac{FWHM} of \SI{1.5}{\mega\hertz} [Fig. \ref{fig:HL}(b)] for \SI{1538.685}{\nano\metre} and \SI{2.8}{\mega\hertz} at \SI{1532.254}{\nano\metre}. The upper bound on the homogeneous linewidth is given by half of the spectral hole linewidth~\cite{moerner1988}, thus leading to a maximum homogeneous linewidth of \SI{0.75}{\mega\hertz} (\SI{3.1}{\nano\electronvolt}) and \SI{1.4}{\mega\hertz} (\SI{5.8}{\nano\electronvolt}) for \SI{1538.685}{\nano\metre} and \SI{1532.254}{\nano\metre}, respectively.

To study the effect of instantaneous spectral diffusion on the spectral hole, the same measurement was carried out using pump and probe widths of \SI{10}{\micro\second} each. 
The measurement was also repeated with a \SI{90}{\micro\second} delay between the pump and probe width. 
The choice of delay was limited by the optical lifetime of the resonance, as a longer delay results in a loss of spectral hole visibility. 
As can be found in Supplemental Material S-E, a delay of \SI{90}{\micro\second} did not affect the hole width and hence instantaneous spectral diffusion does not play a significant role in the spectral hole linewidth on this timescale.

\subsection*{Lifetime}

%Be explicit about background being wavelength independent, conclusion: fairly similar lifetimes
\begin{figure}
	\centering
    \includegraphics[width=1.0\columnwidth]{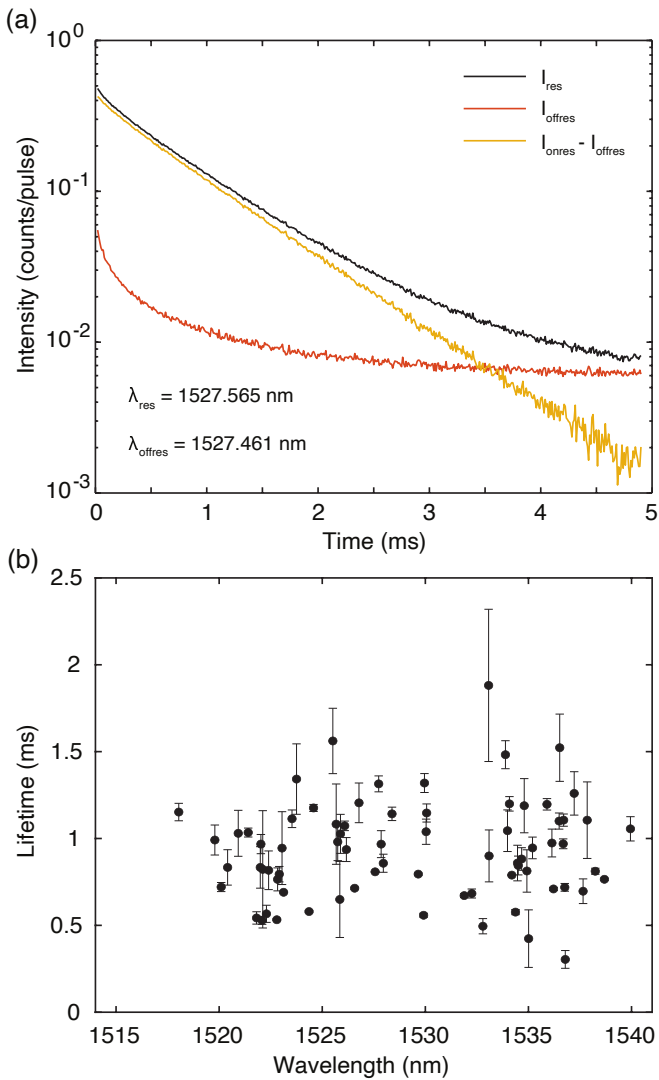}
    \caption{(a) Photoluminescence decay curve for the resonance at \SI{1527.565}{\nano\metre} and the background at \SI{1527.461}{\nano\metre}. 
    (b) Fit result of the lifetime for \num{70} resonances. Errorbars show the standard error of the fit.
    }
    \label{fig:lifetime}
\end{figure}

In the following measurement, the optical lifetime of the \num{70} resonances was measured.
The lifetime of \ce{Er} varies according to the radiative lifetime of the transition as well as the background doping of silicon~\cite{Priolo1998}. 
In addition, in this type of fluorescence measurement, the lifetime measured is typically that of the lowest \ce{^4I_{13/2}} level, regardless of which crystal field level was excited, because higher crystal field levels rapidly decay non-radiatively to the lowest state~\cite{Huang2001}.
In principle, this means that lines with identical fluorescence lifetimes should be associated with the same site.

To measure the fluorescence lifetime, the number of single photon events was repeatedly recorded for \SI{5}{\milli\second} after a \SI{100}{\micro\second} excitation pulse centered on the inhomogeneous peak. 
The signal decay after the excitation pulse did not follow a single exponential curve, but instead was a combination of the resonant signal and a background signal. 
To separate this background decay, it was measured at an optical frequency where the photoluminescence resonance was absent. It was found to be wavelength independent and consists of a fast (\SI{\sim 200}{\micro\second}) and a slow (\SI{\sim 800}{\micro\second}) exponential decay [Fig. \ref{fig:lifetime}(a)]. 
This biexponential decay photoluminescence has been observed in other \ce{Er}:\ce{SI} systems in the past, in \acs{PL} experiments \cite{Coffa1994,Priolo1995,Wu1997,Vinh2005} as well as electroluminescence experiments \cite{Palm1996}. These components are related to the radiative decay of excitons and indirectly excited \ce{Er} ions, respectively. 
The background is thus attributed to a non-competitive process and can be subtracted from the resonance decay signal. 
To get a better estimate on the lifetimes, a nearby background decay approximately two times the \ac{FWHM} at the side was measured at every resonance [Supplemental Material S-D]. 
The difference between the two followed a single exponential decay for most of the resonances, where the largest amplitude decay trace is shown in Fig. \ref{fig:lifetime}(a). 
A number of resonances show a biexponential decay, marked by $*$ in Table \ref{tab:resonances} and presented in the Supplemental Material S-D.

A summary of the lifetimes is presented in Table \ref{tab:resonances}. The optical lifetimes ranged from \SI{0.5}{\milli\second} to \SI{1.5}{\milli\second}, similar to lifetimes found in various \acs{PL} and \acs{PLE} experiments~\cite{Coffa1994,Wu1997,Priolo1998,Taguchi1998,weiss20arXiv} as well as what is expected on theoretical grounds for magnetic dipole transitions~\cite{dodson12}. 
Compared to the \acs{PLE} results in Ref. \cite{weiss20arXiv}, for the seven resonances that are found in both experiments we found a shorter lifetime for each resonance, with the average difference being \SI{250}{\micro\second}.
This reduction in lifetime can be explained by an extra non-radiative mechanism in our crystal, activated by the different geometry or sample parameters.

The distribution of lifetime fits and standard errors [Fig. \ref{fig:lifetime}(b)] does not show separated groups with equal lifetimes, thus it is difficult to associate different peaks within the same site based on the lifetime alone. 

%\begin{figure}
%	\centering
%    \includegraphics[scale=0.7]{FWHMvsPower.png}
%    \caption{FWHM of the homogeneous broadening as function of power on the sample.}
%\end{figure}

% Thompson and friends were looking for sites with (or without) inversion symmetry in a variety of host crystals to find sites with no Stark effect to get less spectral diffusion from charge noise. Might be worth to mention that we should look for those.

%Mention how many ions we detect looking at the count rate

%Studied in detail er in si with resonant pl in contrast to broad pl, sets up for other future studies
%Be more specific about densities we can go to <- to study sites better
%technique is a good way of figuring out which sites we have
%technique applicable other thin films than si, homogeneity, LiNbO,

\section*{Discussion}

In photonic cavities, the maximum Purcell factor $F = 2g/(\gamma_\text{bulk})$ is determined by the optical transition-cavity mode coupling strength $g$ and is achieved for the cavity mode damping rate, i.e. at the transition between weak and strong coupling regimes~\cite{yamaguchi2008}. Because of the measured sub-megahertz dephasing ($\gamma_\text{D}$) and emission rates ($\gamma_\text{sp}$) of the measured \ce{Er} transitions, the damping rate can be reduced to  $\kappa = 4g$. Consequently, the maximum Purcell enhancement is solely defined by the cavity design and the corresponding cavity mode volume $V_\text{m} = 3\lambda^2c/(2\pi n^3F^2\gamma_\text{bulk})$. The \num{6} orders of magnitude Purcell factor required to match state-of-the-art single photon brightness\cite{weiss21} sets an upper bound on the cavity mode volume of $V_\text{m} = 0.1(\lambda/n)^3$ assuming \ce{Er} spontaneous emission rate in \ce{Si} $\gamma_\text{bulk} = \SI{1}{\kilo\hertz}$. Such an ultrasmall cavity mode volume can be achieved in nanobeam \ce{Si} photonic cavities\cite{zhou2019,miura2014} that also provide the required quality factor $Q = 10^5$ ($\kappa = \SI{2}{\giga\hertz}$). The homogeneous linewidth of an \ce{Er} in a nanobeam cavity will be solely limited by the radiative broadening and will exceed the dephasing rate by \num{3} orders of magnitude making \ce{Er} in \ce{Si} a promising platform for carving single photon sources with state-of-the-art brightness and record single photon indistinguishability. 

Various sample parameters can be investigated and optimized such as the varying the \ce{O} and \ce{Er} concentration, isotopically purifying \ce{Si} and fine-tuning the annealing processes for the purpose of narrowing linewidths, relating to long coherence times. 
By reducing the \ce{O} concentration for future samples, the \ce{Er}-only sites can be extracted.
These sites reside in higher symmetry sites since the lattice is undisturbed by the \ce{O}. Likewise, for \ce{Er}-\ce{O} complexes, the inhomogeneity in the environment is affected at different \ce{O} concentrations and thus could alter the linewidths.

The current \ac{SNR} of the \ac{PLE} spectrum suggests that the \ce{Er} concentration can be diminished by up to two orders of magnitude without increasing the repetition number to detect the resonances. 
A reduced ion concentration leads to less ion-ion interactions, potentially narrowing the homogeneous linewidth and increasing the coherence times. 
Another realization to reduce inhomogeneities in the environment is to use isotopically enriched $^{28}$\ce{Si} and by optimizing the annealing procedure. 

\section*{Conclusion}

In this paper we use a novel in situ \ac{PLE} method to study in detail the inhomogeneous linewidths and lifetimes of \num{70} resonances, \num{7} of which have been observed in Ref. \cite{weiss21} and \num{63} which have not been resolved in \ac{EPR}, \ac{PL} and \ac{PLE} experiments before. 
The homogeneous linewidth of a resonance at \SI{1538.685}{\nano\metre} and at \SI{1532.254}{\nano\metre} were studied using transient spectral hole burning and we extracted a homogeneous linewidth of \SI{0.75}{\mega\hertz} and \SI{1.4}{\mega\hertz}, respectively.

The technique presented \SI{300}{\milli\kelvin} allows the characterization of samples on a relatively short timescale and thus measuring the dependence on the parameters discussed previously is an accessible realization. 
Finally, this in situ method can also be applied to optically active dopants in other thin transparent films such as \ce{LiNbO_3} and \ce{SiO_2}. 

%We propose an accessible method to detect multiple \ce{Er} transitions in \ce{Si} with high collection efficiency. The identification and characterization of the transitions which can be detected optically, electrically or both ways is a central step in implementing systems utilizing \ce{Er} in \ce{Si} for quantum communication applications. Additionally, the observation of the multitude of \ce{Er} lines in \ce{Si} with narrow inhomogeneous and homogeneous linewidths provides prospects for using this particular material system in future quantum communication systems. Lastly, the ability to split the inhomogeneous resonances in an applied magnetic field confirms that \ce{Er} ions reside in well defined sites and enables determination of the paramagnetic properties which assists in the identification of the symmetry and structure of the sites.  

\begin{acknowledgments}           
We acknowledge the AFAiiR node of the NCRIS Heavy Ion Capability for access to  
ion-implantation facilities. 
This work was supported by the ARC Centre of Excellence for Quantum Computation and Communication Technology (Grant CE170100012) and the Discovery Project (Grant DP150103699). 
We thank Dr. Sae Woo Nam for kind support to establish the \ac{SSPD} fabrication process at the University of New South Wales, Sydney. 
\end{acknowledgments}  

%\newpage
\section*{Appendix}
\beginappendix
\subsection{Detected Er resonances}
%\newpage

\begin{longtable}{
>{\RaggedRight}p{0.24\columnwidth} >{\RaggedRight}p{0.24\columnwidth} >{\RaggedRight}p{0.24\columnwidth} >{\RaggedRight}p{0.24\columnwidth}}
    \caption{Overview table}\label{tab:resonances}\\
        \toprule
         \thead{Wavelength\\(\si{\nano\metre})}& \thead{Linewidth\\(\si{\giga\hertz})}&
         \thead{Lifetime\\(\si{\milli\second})}&
         \thead{Amplitude\\(counts/pulse)}\\

         \midrule
         \num{1539.949} &\num{2.82} & \num{1.05}  & \num{0.59}\\
         \num{1538.685}$^\dagger$ &\num{1.02} &\num{0.764}  & \num{4.06}\\
         \num{1538.242} &\num{1.81}&\num{0.810}  & \num{1.73}\\
         \num{1537.851} &\num{3.5}&\num{1.1}  & \num{0.26}\\
         \num{1537.652} &\num{1.95}&\num{0.70}  & \num{0.57}\\
         \num{1537.220} &\num{2.5}&\num{1.26}  & \num{0.31}\\
         \num{1536.762} &\num{0.40}&\num{0.72}  & \num{1.29}\\
         \num{1536.708} &\num{0.94}&\num{1.11}  & \num{0.84}\\
         \num{1536.687} &\num{1.5}&\num{0.97}  & \num{1.13}\\
         \num{1536.518} &\num{0.5}&\num{1.52}  & \num{0.25}\\
         \num{1536.489} &\num{1.30}&\num{1.10}  & \num{0.95}\\
         \num{1536.215}$^{*\dagger}$ &\num{1.64}&\num{0.708}  & \num{2.39}\\
         \num{1536.137} &\num{1.7}&\num{0.97}  & \num{0.61}\\
         \num{1535.899} &\num{0.83}&\num{1.19}  & \num{1.40}\\
         \num{1535.199} &\num{0.34}&\num{0.94}  & \num{0.45}\\
%         \num{1535.007} &\num{1.827} & \acs{SNR} too low  &\\
         \num{1534.924} &\num{4.8}&\num{0.81}  & \num{0.25}\\
         \num{1534.796} &\num{0.42}&\num{1.19} & \num{0.27}\\
         \num{1534.672} &\num{4.7}&\num{0.88}  & \num{0.57}\\
         \num{1534.506} &\num{1.3}&\num{0.84}  & \num{0.36}\\
         \num{1534.469} &\num{3.2}&\num{0.86}  & \num{0.31}\\
         \num{1534.371}$^\dagger$ &\num{2.17}&\num{0.575}  & \num{1.00}\\
         \num{1534.195}$^{*\dagger}$ &\num{1.76}&\num{0.788}  & \num{15.96}\\
         \num{1534.080} &\num{1.26}&\num{1.20}  & \num{1.41}\\
         \num{1533.985} &\num{2.3}&\num{1.04}  & \num{0.30}\\
         \num{1533.885} &\num{2.08}&\num{1.48}  & \num{0.80}\\
         \num{1533.087} &\num{3.2}&\num{0.90}  & \num{0.30}\\
%         \num{1533.069} &\num{0.757} & \acs{SNR} too low  &\\
         \num{1532.792}$^\dagger$ &\num{1.6}&\num{0.49}  & \num{0.41}\\
%         \num{1532.371} &\num{2.146} &\num{0.015}  &\\
         \num{1532.254}$^{*\dagger}$ &\num{1.208}&\num{0.682}  & \num{10.63}\\
         \num{1531.886} &\num{0.47}&\num{0.670}  & \num{1.83}\\
         \caption[]{(continued)}\\
         \toprule
         \thead{Wavelength\\(\si{\nano\metre})}& \thead{Linewidth\\(\si{\giga\hertz})}&\thead{Lifetime\\(\si{\milli\second})}&\thead{Amplitude\\(counts/pulse)}\\
         \midrule
         \num{1530.062} &\num{1.69}&\num{1.15}  & \num{0.77}\\
         \num{1530.034} &\num{4.5}&\num{1.04}  & \num{0.78}\\
         \num{1529.955} &\num{3.5}&\num{1.32}  & \num{0.95}\\
         \num{1529.916}$^{*\dagger}$ &\num{0.97}&\num{0.557}  & \num{1.61}\\
         \num{1529.657} &\num{1.68}&\num{0.793}  & \num{3.31}\\
         \num{1528.380} &\num{3.9}&\num{1.14}  & \num{1.53}\\
         \num{1527.963} &\num{3.9}&\num{0.86}  & \num{0.69}\\
         \num{1527.851} &\num{6.9}&\num{0.97}  & \num{0.41}\\
         \num{1527.735} &\num{3.11}&\num{1.31}  & \num{1.06}\\
         \num{1527.565} &\num{1.43}&\num{0.807}  & \num{25.0}\\
         \num{1526.776} &\num{4.9}&\num{1.20}  & \num{0.52}\\
         \num{1526.572}$^*$ &\num{1.70}&\num{0.712}  & \num{3.84}\\
         \num{1526.171} &\num{1.74}&\num{0.93}  & \num{0.69}\\
         \num{1526.088} &\num{1.86}&\num{1.07}  & \num{1.75}\\
         \num{1525.885} &\num{1.8}&\num{1.03}  & \num{0.24}\\
         \num{1525.848} &\num{3.5}&\num{0.6}  & \num{0.15}\\
         \num{1525.751} &\num{2.9}&\num{0.98}  & \num{0.29}\\
         \num{1525.677} &\num{1.4}&\num{1.1}  & \num{0.28}\\
         \num{1525.513} &\num{2.8}&\num{1.56}  & \num{0.32}\\
         \num{1524.577} &\num{2.32}&\num{1.17}  & \num{3.18}\\
         \num{1524.360}$^*$ &\num{1.09}&\num{0.578}  & \num{2.41}\\
         \num{1523.753} &\num{3.2}&\num{1.3} & \num{0.39}\\
         \num{1523.535} &\num{3.8}&\num{1.11}  & \num{0.85}\\
         \num{1523.126}$^*$ &\num{2.07}&\num{0.689}  & \num{7.45}\\
         \num{1523.050} &\num{2.9}&\num{0.9}  & \num{0.27}\\
         \num{1522.917} &\num{2.6}&\num{0.79}  & \num{0.81}\\
         \num{1522.835} &\num{1.1}&\num{0.76}  & \num{0.49}\\
         \num{1522.797}$^*$ &\num{0.74}&\num{0.531}  & \num{2.61}\\
         \num{1522.399} &\num{1.0}&\num{0.82}  & \num{0.24}\\
         \num{1522.291} &\num{3.6}&\num{0.57}  & \num{0.39}\\
         \num{1522.114} &\num{1.0}&\num{0.8}  & \num{0.19}\\
         \num{1522.085} &\num{1.18}&\num{0.53}  & \num{0.63}\\
         \num{1522.025} &\num{1.13}&\num{0.97}  & \num{0.48}\\
         \num{1521.994} &\num{1.2}&\num{0.83}  & \num{0.26}\\
         \num{1521.816} &\num{2.7}&\num{0.54}  & \num{0.45}\\
         \num{1521.409} &\num{0.65}&\num{1.03}  & \num{1.25}\\
         \num{1520.926} &\num{3.6}&\num{1.03}  & \num{0.24}\\
         \num{1520.412} &\num{2.5}&\num{0.83}  & \num{0.35}\\
         \num{1520.094} &\num{9.1}&\num{0.72}  & \num{0.86}\\
         \num{1519.793} &\num{2.2}&\num{0.99}  & \num{0.56}\\
         \num{1518.042} &\num{5.3}&\num{1.15}  & \num{0.61}\\
         
        \bottomrule
\end{longtable}

\noindent$\dagger$ Observed in Ref. \cite{weiss20arXiv}.\\
\noindent* Biexponential decay.

\bibliographystyle{apsrev4-2}
\bibliography{bibliography}

\begin{thebibliography}{49}%
\makeatletter
\providecommand \@ifxundefined [1]{%
 \@ifx{#1\undefined}
}%
\providecommand \@ifnum [1]{%
 \ifnum #1\expandafter \@firstoftwo
 \else \expandafter \@secondoftwo
 \fi
}%
\providecommand \@ifx [1]{%
 \ifx #1\expandafter \@firstoftwo
 \else \expandafter \@secondoftwo
 \fi
}%
\providecommand \natexlab [1]{#1}%
\providecommand \enquote  [1]{``#1''}%
\providecommand \bibnamefont  [1]{#1}%
\providecommand \bibfnamefont [1]{#1}%
\providecommand \citenamefont [1]{#1}%
\providecommand \href@noop [0]{\@secondoftwo}%
\providecommand \href [0]{\begingroup \@sanitize@url \@href}%
\providecommand \@href[1]{\@@startlink{#1}\@@href}%
\providecommand \@@href[1]{\endgroup#1\@@endlink}%
\providecommand \@sanitize@url [0]{\catcode `\\12\catcode `\$12\catcode
  `\&12\catcode `\#12\catcode `\^12\catcode `\_12\catcode `\%12\relax}%
\providecommand \@@startlink[1]{}%
\providecommand \@@endlink[0]{}%
\providecommand \url  [0]{\begingroup\@sanitize@url \@url }%
\providecommand \@url [1]{\endgroup\@href {#1}{\urlprefix }}%
\providecommand \urlprefix  [0]{URL }%
\providecommand \Eprint [0]{\href }%
\providecommand \doibase [0]{https://doi.org/}%
\providecommand \selectlanguage [0]{\@gobble}%
\providecommand \bibinfo  [0]{\@secondoftwo}%
\providecommand \bibfield  [0]{\@secondoftwo}%
\providecommand \translation [1]{[#1]}%
\providecommand \BibitemOpen [0]{}%
\providecommand \bibitemStop [0]{}%
\providecommand \bibitemNoStop [0]{.\EOS\space}%
\providecommand \EOS [0]{\spacefactor3000\relax}%
\providecommand \BibitemShut  [1]{\csname bibitem#1\endcsname}%
\let\auto@bib@innerbib\@empty
%</preamble>
\bibitem [{\citenamefont {Thiel}\ \emph {et~al.}(2011)\citenamefont {Thiel},
  \citenamefont {B\"ottger},\ and\ \citenamefont {Cone}}]{thiel2011}%
  \BibitemOpen
  \bibfield  {author} {\bibinfo {author} {\bibfnamefont {C.~W.}\ \bibnamefont
  {Thiel}}, \bibinfo {author} {\bibfnamefont {T.}~\bibnamefont {B\"ottger}},\
  and\ \bibinfo {author} {\bibfnamefont {R.~L.}\ \bibnamefont {Cone}},\ }\href
  {https://doi.org/10.1016/j.jlumin.2010.12.015} {\bibfield  {journal}
  {\bibinfo  {journal} {Journal of Luminescence}\ }\textbf {\bibinfo {volume}
  {131}},\ \bibinfo {pages} {353} (\bibinfo {year} {2011})}\BibitemShut
  {NoStop}%
\bibitem [{\citenamefont {B\"ottger}\ \emph {et~al.}(2009)\citenamefont
  {B\"ottger}, \citenamefont {Thiel}, \citenamefont {Cone},\ and\ \citenamefont
  {Sun}}]{Bottger2009}%
  \BibitemOpen
  \bibfield  {author} {\bibinfo {author} {\bibfnamefont {T.}~\bibnamefont
  {B\"ottger}}, \bibinfo {author} {\bibfnamefont {C.~W.}\ \bibnamefont
  {Thiel}}, \bibinfo {author} {\bibfnamefont {R.~L.}\ \bibnamefont {Cone}},\
  and\ \bibinfo {author} {\bibfnamefont {Y.}~\bibnamefont {Sun}},\ }\bibfield
  {journal} {\bibinfo  {journal} {Phys. Rev. B}\ }\textbf {\bibinfo {volume}
  {79}},\ \href {https://doi.org/10.1103/PhysRevB.79.115104}
  {10.1103/PhysRevB.79.115104} (\bibinfo {year} {2009})\BibitemShut {NoStop}%
\bibitem [{\citenamefont {Zhong}\ \emph {et~al.}(2015)\citenamefont {Zhong},
  \citenamefont {Hedges}, \citenamefont {Ahlefeldt}, \citenamefont
  {Bartholomew}, \citenamefont {Beavan}, \citenamefont {Wittig}, \citenamefont
  {Longdell},\ and\ \citenamefont {Sellars}}]{Zhong2015}%
  \BibitemOpen
  \bibfield  {author} {\bibinfo {author} {\bibfnamefont {M.~J.}\ \bibnamefont
  {Zhong}}, \bibinfo {author} {\bibfnamefont {M.~P.}\ \bibnamefont {Hedges}},
  \bibinfo {author} {\bibfnamefont {R.~L.}\ \bibnamefont {Ahlefeldt}}, \bibinfo
  {author} {\bibfnamefont {J.~G.}\ \bibnamefont {Bartholomew}}, \bibinfo
  {author} {\bibfnamefont {S.~E.}\ \bibnamefont {Beavan}}, \bibinfo {author}
  {\bibfnamefont {S.~M.}\ \bibnamefont {Wittig}}, \bibinfo {author}
  {\bibfnamefont {J.~J.}\ \bibnamefont {Longdell}},\ and\ \bibinfo {author}
  {\bibfnamefont {M.~J.}\ \bibnamefont {Sellars}},\ }\href
  {https://doi.org/10.1038/nature14025} {\bibfield  {journal} {\bibinfo
  {journal} {Nature}\ }\textbf {\bibinfo {volume} {517}},\ \bibinfo {pages}
  {177} (\bibinfo {year} {2015})}\BibitemShut {NoStop}%
\bibitem [{\citenamefont {Ran\v{c}i\'{c}}\ \emph {et~al.}(2017)\citenamefont
  {Ran\v{c}i\'{c}}, \citenamefont {Hedges}, \citenamefont {Ahlefeldt},\ and\
  \citenamefont {Sellars}}]{Rancic2017}%
  \BibitemOpen
  \bibfield  {author} {\bibinfo {author} {\bibfnamefont {M.}~\bibnamefont
  {Ran\v{c}i\'{c}}}, \bibinfo {author} {\bibfnamefont {M.~P.}\ \bibnamefont
  {Hedges}}, \bibinfo {author} {\bibfnamefont {R.~L.}\ \bibnamefont
  {Ahlefeldt}},\ and\ \bibinfo {author} {\bibfnamefont {M.~J.}\ \bibnamefont
  {Sellars}},\ }\href {https://doi.org/10.1038/nphys4254} {\bibfield  {journal}
  {\bibinfo  {journal} {Nature Physics}\ }\textbf {\bibinfo {volume} {14}},\
  \bibinfo {pages} {50} (\bibinfo {year} {2017})}\BibitemShut {NoStop}%
\bibitem [{\citenamefont {Kolesov}\ \emph {et~al.}(2012)\citenamefont
  {Kolesov}, \citenamefont {Xia}, \citenamefont {Reuter}, \citenamefont
  {Stohr}, \citenamefont {Zappe}, \citenamefont {Meijer}, \citenamefont
  {Hemmer},\ and\ \citenamefont {Wrachtrup}}]{kolesov2012}%
  \BibitemOpen
  \bibfield  {author} {\bibinfo {author} {\bibfnamefont {R.}~\bibnamefont
  {Kolesov}}, \bibinfo {author} {\bibfnamefont {K.}~\bibnamefont {Xia}},
  \bibinfo {author} {\bibfnamefont {R.}~\bibnamefont {Reuter}}, \bibinfo
  {author} {\bibfnamefont {R.}~\bibnamefont {Stohr}}, \bibinfo {author}
  {\bibfnamefont {A.}~\bibnamefont {Zappe}}, \bibinfo {author} {\bibfnamefont
  {J.}~\bibnamefont {Meijer}}, \bibinfo {author} {\bibfnamefont {P.~R.}\
  \bibnamefont {Hemmer}},\ and\ \bibinfo {author} {\bibfnamefont
  {J.}~\bibnamefont {Wrachtrup}},\ }\bibfield  {journal} {\bibinfo  {journal}
  {Nature Communications}\ }\textbf {\bibinfo {volume} {3}},\ \href
  {https://doi.org/10.1038/ncomms2034} {10.1038/ncomms2034} (\bibinfo {year}
  {2012})\BibitemShut {NoStop}%
\bibitem [{\citenamefont {Siyushev}\ \emph {et~al.}(2014)\citenamefont
  {Siyushev}, \citenamefont {Xia}, \citenamefont {Reuter}, \citenamefont
  {Jamali}, \citenamefont {Zhao}, \citenamefont {Yang}, \citenamefont {Duan},
  \citenamefont {Kukharchyk}, \citenamefont {Wieck}, \citenamefont {Kolesov},\
  and\ \citenamefont {Wrachtrup}}]{siyushev2014}%
  \BibitemOpen
  \bibfield  {author} {\bibinfo {author} {\bibfnamefont {P.}~\bibnamefont
  {Siyushev}}, \bibinfo {author} {\bibfnamefont {K.}~\bibnamefont {Xia}},
  \bibinfo {author} {\bibfnamefont {R.}~\bibnamefont {Reuter}}, \bibinfo
  {author} {\bibfnamefont {M.}~\bibnamefont {Jamali}}, \bibinfo {author}
  {\bibfnamefont {N.}~\bibnamefont {Zhao}}, \bibinfo {author} {\bibfnamefont
  {N.}~\bibnamefont {Yang}}, \bibinfo {author} {\bibfnamefont {C.}~\bibnamefont
  {Duan}}, \bibinfo {author} {\bibfnamefont {N.}~\bibnamefont {Kukharchyk}},
  \bibinfo {author} {\bibfnamefont {A.~D.}\ \bibnamefont {Wieck}}, \bibinfo
  {author} {\bibfnamefont {R.}~\bibnamefont {Kolesov}},\ and\ \bibinfo {author}
  {\bibfnamefont {J.}~\bibnamefont {Wrachtrup}},\ }\bibfield  {journal}
  {\bibinfo  {journal} {Nature Communications}\ }\textbf {\bibinfo {volume}
  {5}},\ \href {https://doi.org/10.1038/ncomms4895} {10.1038/ncomms4895}
  (\bibinfo {year} {2014})\BibitemShut {NoStop}%
\bibitem [{\citenamefont {Kindem}\ \emph {et~al.}(2020)\citenamefont {Kindem},
  \citenamefont {Ruskuc}, \citenamefont {Bartholomew}, \citenamefont {Rochman},
  \citenamefont {Huan},\ and\ \citenamefont {Faraon}}]{kindem2020}%
  \BibitemOpen
  \bibfield  {author} {\bibinfo {author} {\bibfnamefont {J.~M.}\ \bibnamefont
  {Kindem}}, \bibinfo {author} {\bibfnamefont {A.}~\bibnamefont {Ruskuc}},
  \bibinfo {author} {\bibfnamefont {J.~G.}\ \bibnamefont {Bartholomew}},
  \bibinfo {author} {\bibfnamefont {J.}~\bibnamefont {Rochman}}, \bibinfo
  {author} {\bibfnamefont {Y.~Q.}\ \bibnamefont {Huan}},\ and\ \bibinfo
  {author} {\bibfnamefont {A.}~\bibnamefont {Faraon}},\ }\href
  {https://doi.org/10.1038/s41586-020-2160-9} {\bibfield  {journal} {\bibinfo
  {journal} {Nature}\ }\textbf {\bibinfo {volume} {580}},\ \bibinfo {pages}
  {201} (\bibinfo {year} {2020})}\BibitemShut {NoStop}%
\bibitem [{\citenamefont {Raha}\ \emph {et~al.}(2020)\citenamefont {Raha},
  \citenamefont {Chen}, \citenamefont {Phenicie}, \citenamefont {Ourari},
  \citenamefont {Dibos},\ and\ \citenamefont {Thompson}}]{raha20}%
  \BibitemOpen
  \bibfield  {author} {\bibinfo {author} {\bibfnamefont {M.}~\bibnamefont
  {Raha}}, \bibinfo {author} {\bibfnamefont {S.}~\bibnamefont {Chen}}, \bibinfo
  {author} {\bibfnamefont {C.~M.}\ \bibnamefont {Phenicie}}, \bibinfo {author}
  {\bibfnamefont {S.}~\bibnamefont {Ourari}}, \bibinfo {author} {\bibfnamefont
  {A.~M.}\ \bibnamefont {Dibos}},\ and\ \bibinfo {author} {\bibfnamefont
  {J.~D.}\ \bibnamefont {Thompson}},\ }\href
  {https://doi.org/10.1038/s41467-020-15138-7} {\bibfield  {journal} {\bibinfo
  {journal} {Nature Communications}\ }\textbf {\bibinfo {volume} {11}},\
  \bibinfo {pages} {1605} (\bibinfo {year} {2020})}\BibitemShut {NoStop}%
\bibitem [{\citenamefont {Chen}\ \emph {et~al.}(2020)\citenamefont {Chen},
  \citenamefont {Raha}, \citenamefont {Phenicie}, \citenamefont {Ourari},\ and\
  \citenamefont {Thompson}}]{chen2020}%
  \BibitemOpen
  \bibfield  {author} {\bibinfo {author} {\bibfnamefont {S.}~\bibnamefont
  {Chen}}, \bibinfo {author} {\bibfnamefont {M.}~\bibnamefont {Raha}}, \bibinfo
  {author} {\bibfnamefont {C.~M.}\ \bibnamefont {Phenicie}}, \bibinfo {author}
  {\bibfnamefont {S.}~\bibnamefont {Ourari}},\ and\ \bibinfo {author}
  {\bibfnamefont {J.~D.}\ \bibnamefont {Thompson}},\ }\href
  {https://doi.org/10.1126/science.abc7821} {\bibfield  {journal} {\bibinfo
  {journal} {Science}\ }\textbf {\bibinfo {volume} {370}},\ \bibinfo {pages}
  {592} (\bibinfo {year} {2020})}\BibitemShut {NoStop}%
\bibitem [{\citenamefont {Utikal}\ \emph {et~al.}(2014)\citenamefont {Utikal},
  \citenamefont {Eichhammer}, \citenamefont {Petersen}, \citenamefont {Renn},
  \citenamefont {Goetzinger},\ and\ \citenamefont {Sandoghdar}}]{utikal2014}%
  \BibitemOpen
  \bibfield  {author} {\bibinfo {author} {\bibfnamefont {T.}~\bibnamefont
  {Utikal}}, \bibinfo {author} {\bibfnamefont {E.}~\bibnamefont {Eichhammer}},
  \bibinfo {author} {\bibfnamefont {L.}~\bibnamefont {Petersen}}, \bibinfo
  {author} {\bibfnamefont {A.}~\bibnamefont {Renn}}, \bibinfo {author}
  {\bibfnamefont {S.}~\bibnamefont {Goetzinger}},\ and\ \bibinfo {author}
  {\bibfnamefont {V.}~\bibnamefont {Sandoghdar}},\ }\bibfield  {journal}
  {\bibinfo  {journal} {Nature Communications}\ }\textbf {\bibinfo {volume}
  {5}},\ \href {https://doi.org/10.1038/ncomms4627} {10.1038/ncomms4627}
  (\bibinfo {year} {2014})\BibitemShut {NoStop}%
\bibitem [{\citenamefont {Yin}\ \emph {et~al.}(2013)\citenamefont {Yin},
  \citenamefont {Ran\v{c}i\'{c}}, \citenamefont {de~Boo}, \citenamefont
  {Stavrias}, \citenamefont {McCallum}, \citenamefont {Sellars},\ and\
  \citenamefont {Rogge}}]{yin13}%
  \BibitemOpen
  \bibfield  {author} {\bibinfo {author} {\bibfnamefont {C.}~\bibnamefont
  {Yin}}, \bibinfo {author} {\bibfnamefont {M.}~\bibnamefont {Ran\v{c}i\'{c}}},
  \bibinfo {author} {\bibfnamefont {G.~G.}\ \bibnamefont {de~Boo}}, \bibinfo
  {author} {\bibfnamefont {N.}~\bibnamefont {Stavrias}}, \bibinfo {author}
  {\bibfnamefont {J.~C.}\ \bibnamefont {McCallum}}, \bibinfo {author}
  {\bibfnamefont {M.~J.}\ \bibnamefont {Sellars}},\ and\ \bibinfo {author}
  {\bibfnamefont {S.}~\bibnamefont {Rogge}},\ }\href
  {https://doi.org/10.1038/nature12081} {\bibfield  {journal} {\bibinfo
  {journal} {Nature}\ }\textbf {\bibinfo {volume} {497}},\ \bibinfo {pages}
  {91} (\bibinfo {year} {2013})}\BibitemShut {NoStop}%
\bibitem [{\citenamefont {Zhong}\ \emph {et~al.}(2017)\citenamefont {Zhong},
  \citenamefont {Kindem}, \citenamefont {Bartholomew}, \citenamefont {Rochman},
  \citenamefont {Craiciu}, \citenamefont {Miyazono}, \citenamefont
  {Bettinelli}, \citenamefont {Cavalli}, \citenamefont {Verma}, \citenamefont
  {Nam}, \citenamefont {Marsili}, \citenamefont {Shaw}, \citenamefont {Beyer},\
  and\ \citenamefont {Faraon}}]{zhong2017}%
  \BibitemOpen
  \bibfield  {author} {\bibinfo {author} {\bibfnamefont {T.}~\bibnamefont
  {Zhong}}, \bibinfo {author} {\bibfnamefont {J.~M.}\ \bibnamefont {Kindem}},
  \bibinfo {author} {\bibfnamefont {J.~G.}\ \bibnamefont {Bartholomew}},
  \bibinfo {author} {\bibfnamefont {J.}~\bibnamefont {Rochman}}, \bibinfo
  {author} {\bibfnamefont {I.}~\bibnamefont {Craiciu}}, \bibinfo {author}
  {\bibfnamefont {E.}~\bibnamefont {Miyazono}}, \bibinfo {author}
  {\bibfnamefont {M.}~\bibnamefont {Bettinelli}}, \bibinfo {author}
  {\bibfnamefont {E.}~\bibnamefont {Cavalli}}, \bibinfo {author} {\bibfnamefont
  {V.}~\bibnamefont {Verma}}, \bibinfo {author} {\bibfnamefont {S.~W.}\
  \bibnamefont {Nam}}, \bibinfo {author} {\bibfnamefont {F.}~\bibnamefont
  {Marsili}}, \bibinfo {author} {\bibfnamefont {M.~D.}\ \bibnamefont {Shaw}},
  \bibinfo {author} {\bibfnamefont {A.~D.}\ \bibnamefont {Beyer}},\ and\
  \bibinfo {author} {\bibfnamefont {A.}~\bibnamefont {Faraon}},\ }\href
  {https://doi.org/10.1126/science.aan5959} {\bibfield  {journal} {\bibinfo
  {journal} {Science}\ }\textbf {\bibinfo {volume} {357}},\ \bibinfo {pages}
  {1392} (\bibinfo {year} {2017})}\BibitemShut {NoStop}%
\bibitem [{\citenamefont {Grimm}\ \emph {et~al.}(2021)\citenamefont {Grimm},
  \citenamefont {Beckert}, \citenamefont {Aeppli},\ and\ \citenamefont
  {M\"uller}}]{grimm2021}%
  \BibitemOpen
  \bibfield  {author} {\bibinfo {author} {\bibfnamefont {M.}~\bibnamefont
  {Grimm}}, \bibinfo {author} {\bibfnamefont {A.}~\bibnamefont {Beckert}},
  \bibinfo {author} {\bibfnamefont {G.}~\bibnamefont {Aeppli}},\ and\ \bibinfo
  {author} {\bibfnamefont {M.}~\bibnamefont {M\"uller}},\ }\href
  {https://doi.org/10.1103/PRXQuantum.2.010312} {\bibfield  {journal} {\bibinfo
   {journal} {PRX Quantum}\ }\textbf {\bibinfo {volume} {2}},\ \bibinfo {pages}
  {010312} (\bibinfo {year} {2021})}\BibitemShut {NoStop}%
\bibitem [{\citenamefont {Kenyon}(2005)}]{kenyon2005}%
  \BibitemOpen
  \bibfield  {author} {\bibinfo {author} {\bibfnamefont {A.~J.}\ \bibnamefont
  {Kenyon}},\ }\href {https://doi.org/10.1088/0268-1242/20/12/R02} {\bibfield
  {journal} {\bibinfo  {journal} {Semiconductor Science and Technology}\
  }\textbf {\bibinfo {volume} {20}},\ \bibinfo {pages} {R65} (\bibinfo {year}
  {2005})}\BibitemShut {NoStop}%
\bibitem [{\citenamefont {Weiss}\ \emph {et~al.}(2021)\citenamefont {Weiss},
  \citenamefont {Gritsch}, \citenamefont {Merkel},\ and\ \citenamefont
  {Reiserer}}]{weiss21}%
  \BibitemOpen
  \bibfield  {author} {\bibinfo {author} {\bibfnamefont {L.}~\bibnamefont
  {Weiss}}, \bibinfo {author} {\bibfnamefont {A.}~\bibnamefont {Gritsch}},
  \bibinfo {author} {\bibfnamefont {B.}~\bibnamefont {Merkel}},\ and\ \bibinfo
  {author} {\bibfnamefont {A.}~\bibnamefont {Reiserer}},\ }\href
  {https://doi.org/10.1364/OPTICA.413330} {\bibfield  {journal} {\bibinfo
  {journal} {Optica}\ }\textbf {\bibinfo {volume} {8}},\ \bibinfo {pages} {40}
  (\bibinfo {year} {2021})}\BibitemShut {NoStop}%
\bibitem [{\citenamefont {Steger}\ \emph {et~al.}(2012)\citenamefont {Steger},
  \citenamefont {Saeedi}, \citenamefont {Thewalt}, \citenamefont {Morton},
  \citenamefont {Riemann}, \citenamefont {Abrosimov}, \citenamefont {Becker},\
  and\ \citenamefont {Pohl}}]{steger2012}%
  \BibitemOpen
  \bibfield  {author} {\bibinfo {author} {\bibfnamefont {M.}~\bibnamefont
  {Steger}}, \bibinfo {author} {\bibfnamefont {K.}~\bibnamefont {Saeedi}},
  \bibinfo {author} {\bibfnamefont {M.~L.~W.}\ \bibnamefont {Thewalt}},
  \bibinfo {author} {\bibfnamefont {J.~J.~L.}\ \bibnamefont {Morton}}, \bibinfo
  {author} {\bibfnamefont {H.}~\bibnamefont {Riemann}}, \bibinfo {author}
  {\bibfnamefont {N.~V.}\ \bibnamefont {Abrosimov}}, \bibinfo {author}
  {\bibfnamefont {P.}~\bibnamefont {Becker}},\ and\ \bibinfo {author}
  {\bibfnamefont {H.~J.}\ \bibnamefont {Pohl}},\ }\href
  {https://doi.org/10.1126/science.1217635} {\bibfield  {journal} {\bibinfo
  {journal} {Science}\ }\textbf {\bibinfo {volume} {336}},\ \bibinfo {pages}
  {1280} (\bibinfo {year} {2012})}\BibitemShut {NoStop}%
\bibitem [{\citenamefont {Muhonen}\ \emph {et~al.}(2014)\citenamefont
  {Muhonen}, \citenamefont {Dehollain}, \citenamefont {Laucht}, \citenamefont
  {Hudson}, \citenamefont {Kalra}, \citenamefont {Sekiguchi}, \citenamefont
  {Itoh}, \citenamefont {Jamieson}, \citenamefont {McCallum}, \citenamefont
  {Dzurak},\ and\ \citenamefont {Morello}}]{muhonen2014}%
  \BibitemOpen
  \bibfield  {author} {\bibinfo {author} {\bibfnamefont {J.~T.}\ \bibnamefont
  {Muhonen}}, \bibinfo {author} {\bibfnamefont {J.~P.}\ \bibnamefont
  {Dehollain}}, \bibinfo {author} {\bibfnamefont {A.}~\bibnamefont {Laucht}},
  \bibinfo {author} {\bibfnamefont {F.~E.}\ \bibnamefont {Hudson}}, \bibinfo
  {author} {\bibfnamefont {R.}~\bibnamefont {Kalra}}, \bibinfo {author}
  {\bibfnamefont {T.}~\bibnamefont {Sekiguchi}}, \bibinfo {author}
  {\bibfnamefont {K.~M.}\ \bibnamefont {Itoh}}, \bibinfo {author}
  {\bibfnamefont {D.~N.}\ \bibnamefont {Jamieson}}, \bibinfo {author}
  {\bibfnamefont {J.~C.}\ \bibnamefont {McCallum}}, \bibinfo {author}
  {\bibfnamefont {A.~S.}\ \bibnamefont {Dzurak}},\ and\ \bibinfo {author}
  {\bibfnamefont {A.}~\bibnamefont {Morello}},\ }\href
  {https://doi.org/10.1038/Nnano.2014.211} {\bibfield  {journal} {\bibinfo
  {journal} {Nature Nanotechnology}\ }\textbf {\bibinfo {volume} {9}},\
  \bibinfo {pages} {986} (\bibinfo {year} {2014})}\BibitemShut {NoStop}%
\bibitem [{\citenamefont {Bergeron}\ \emph {et~al.}(2020)\citenamefont
  {Bergeron}, \citenamefont {Chartrand}, \citenamefont {Kurkjian},
  \citenamefont {Morse}, \citenamefont {Riemann}, \citenamefont {Abrosimov},
  \citenamefont {Becker}, \citenamefont {Pohl}, \citenamefont {Thewalt},\ and\
  \citenamefont {Simmons}}]{bergeron2020}%
  \BibitemOpen
  \bibfield  {author} {\bibinfo {author} {\bibfnamefont {L.}~\bibnamefont
  {Bergeron}}, \bibinfo {author} {\bibfnamefont {C.}~\bibnamefont {Chartrand}},
  \bibinfo {author} {\bibfnamefont {A.~T.~K.}\ \bibnamefont {Kurkjian}},
  \bibinfo {author} {\bibfnamefont {K.~J.}\ \bibnamefont {Morse}}, \bibinfo
  {author} {\bibfnamefont {H.}~\bibnamefont {Riemann}}, \bibinfo {author}
  {\bibfnamefont {N.~V.}\ \bibnamefont {Abrosimov}}, \bibinfo {author}
  {\bibfnamefont {P.}~\bibnamefont {Becker}}, \bibinfo {author} {\bibfnamefont
  {H.~J.}\ \bibnamefont {Pohl}}, \bibinfo {author} {\bibfnamefont {M.~L.~W.}\
  \bibnamefont {Thewalt}},\ and\ \bibinfo {author} {\bibfnamefont
  {S.}~\bibnamefont {Simmons}},\ }\href
  {https://doi.org/10.1103/PRXQuantum.1.020301} {\bibfield  {journal} {\bibinfo
   {journal} {PRX Quantum}\ }\textbf {\bibinfo {volume} {1}},\ \bibinfo {pages}
  {020301} (\bibinfo {year} {2020})}\BibitemShut {NoStop}%
\bibitem [{\citenamefont {Yang}\ \emph {et~al.}(2009)\citenamefont {Yang},
  \citenamefont {Steger}, \citenamefont {Sekiguchi}, \citenamefont {Thewalt},
  \citenamefont {Ager},\ and\ \citenamefont {Haller}}]{yang2009}%
  \BibitemOpen
  \bibfield  {author} {\bibinfo {author} {\bibfnamefont {A.}~\bibnamefont
  {Yang}}, \bibinfo {author} {\bibfnamefont {M.}~\bibnamefont {Steger}},
  \bibinfo {author} {\bibfnamefont {T.}~\bibnamefont {Sekiguchi}}, \bibinfo
  {author} {\bibfnamefont {M.~L.~W.}\ \bibnamefont {Thewalt}}, \bibinfo
  {author} {\bibfnamefont {J.~W.}\ \bibnamefont {Ager}},\ and\ \bibinfo
  {author} {\bibfnamefont {E.~E.}\ \bibnamefont {Haller}},\ }\bibfield
  {journal} {\bibinfo  {journal} {Applied Physics Letters}\ }\textbf {\bibinfo
  {volume} {95}},\ \href {https://doi.org/10.1063/1.3238268}
  {10.1063/1.3238268} (\bibinfo {year} {2009})\BibitemShut {NoStop}%
\bibitem [{\citenamefont {Przybyli\ifmmode~\acute{n}\else \'{n}\fi{}ska}\ \emph
  {et~al.}(1996)\citenamefont {Przybyli\ifmmode~\acute{n}\else \'{n}\fi{}ska},
  \citenamefont {Jantsch}, \citenamefont {Suprun-Belevitch}, \citenamefont
  {Stepikhova}, \citenamefont {Palmetshofer}, \citenamefont {Hendorfer},
  \citenamefont {Kozanecki}, \citenamefont {Wilson},\ and\ \citenamefont
  {Sealy}}]{przybylinska1996}%
  \BibitemOpen
  \bibfield  {author} {\bibinfo {author} {\bibfnamefont {H.}~\bibnamefont
  {Przybyli\ifmmode~\acute{n}\else \'{n}\fi{}ska}}, \bibinfo {author}
  {\bibfnamefont {W.}~\bibnamefont {Jantsch}}, \bibinfo {author} {\bibfnamefont
  {Y.}~\bibnamefont {Suprun-Belevitch}}, \bibinfo {author} {\bibfnamefont
  {M.}~\bibnamefont {Stepikhova}}, \bibinfo {author} {\bibfnamefont
  {L.}~\bibnamefont {Palmetshofer}}, \bibinfo {author} {\bibfnamefont
  {G.}~\bibnamefont {Hendorfer}}, \bibinfo {author} {\bibfnamefont
  {A.}~\bibnamefont {Kozanecki}}, \bibinfo {author} {\bibfnamefont {R.~J.}\
  \bibnamefont {Wilson}},\ and\ \bibinfo {author} {\bibfnamefont {B.~J.}\
  \bibnamefont {Sealy}},\ }\href {https://doi.org/10.1103/PhysRevB.54.2532}
  {\bibfield  {journal} {\bibinfo  {journal} {Physical Review B}\ }\textbf
  {\bibinfo {volume} {54}},\ \bibinfo {pages} {2532} (\bibinfo {year}
  {1996})}\BibitemShut {NoStop}%
\bibitem [{\citenamefont {Hogg}\ \emph {et~al.}(1996)\citenamefont {Hogg},
  \citenamefont {Takahei},\ and\ \citenamefont {Taguchi}}]{hogg1996}%
  \BibitemOpen
  \bibfield  {author} {\bibinfo {author} {\bibfnamefont {R.~A.}\ \bibnamefont
  {Hogg}}, \bibinfo {author} {\bibfnamefont {K.}~\bibnamefont {Takahei}},\ and\
  \bibinfo {author} {\bibfnamefont {A.}~\bibnamefont {Taguchi}},\ }\href
  {https://doi.org/10.1063/1.362494} {\bibfield  {journal} {\bibinfo  {journal}
  {Journal of Applied Physics}\ }\textbf {\bibinfo {volume} {79}},\ \bibinfo
  {pages} {8682} (\bibinfo {year} {1996})}\BibitemShut {NoStop}%
\bibitem [{\citenamefont {Palm}\ \emph {et~al.}(1996)\citenamefont {Palm},
  \citenamefont {Gan}, \citenamefont {Zheng}, \citenamefont {Michel},\ and\
  \citenamefont {Kimerling}}]{Palm1996}%
  \BibitemOpen
  \bibfield  {author} {\bibinfo {author} {\bibfnamefont {J.}~\bibnamefont
  {Palm}}, \bibinfo {author} {\bibfnamefont {F.}~\bibnamefont {Gan}}, \bibinfo
  {author} {\bibfnamefont {B.}~\bibnamefont {Zheng}}, \bibinfo {author}
  {\bibfnamefont {J.}~\bibnamefont {Michel}},\ and\ \bibinfo {author}
  {\bibfnamefont {L.~C.}\ \bibnamefont {Kimerling}},\ }\href
  {https://doi.org/10.1103/physrevb.54.17603} {\bibfield  {journal} {\bibinfo
  {journal} {Phys Rev B Condens Matter}\ }\textbf {\bibinfo {volume} {54}},\
  \bibinfo {pages} {17603} (\bibinfo {year} {1996})}\BibitemShut {NoStop}%
\bibitem [{\citenamefont {Takahei}\ and\ \citenamefont
  {Taguchi}(1995)}]{takahei1995}%
  \BibitemOpen
  \bibfield  {author} {\bibinfo {author} {\bibfnamefont {K.}~\bibnamefont
  {Takahei}}\ and\ \bibinfo {author} {\bibfnamefont {A.}~\bibnamefont
  {Taguchi}},\ }\href {https://doi.org/10.1063/1.358866} {\bibfield  {journal}
  {\bibinfo  {journal} {Journal of Applied Physics}\ }\textbf {\bibinfo
  {volume} {77}},\ \bibinfo {pages} {1735} (\bibinfo {year}
  {1995})}\BibitemShut {NoStop}%
\bibitem [{\citenamefont {Weiss}\ \emph {et~al.}(2020)\citenamefont {Weiss},
  \citenamefont {Gritsch}, \citenamefont {Merkel},\ and\ \citenamefont
  {Reiserer}}]{weiss20arXiv}%
  \BibitemOpen
  \bibfield  {author} {\bibinfo {author} {\bibfnamefont {L.}~\bibnamefont
  {Weiss}}, \bibinfo {author} {\bibfnamefont {A.}~\bibnamefont {Gritsch}},
  \bibinfo {author} {\bibfnamefont {B.}~\bibnamefont {Merkel}},\ and\ \bibinfo
  {author} {\bibfnamefont {A.}~\bibnamefont {Reiserer}},\ }\href
  {https://arxiv.org/abs/2005.01775} {\bibfield  {journal} {\bibinfo  {journal}
  {ArXiv}\ } (\bibinfo {year} {2020})}\BibitemShut {NoStop}%
\bibitem [{\citenamefont {de~Boo}\ \emph {et~al.}(2020)\citenamefont {de~Boo},
  \citenamefont {Yin}, \citenamefont {Ran\v{c}i\'{c}}, \citenamefont {Johnson},
  \citenamefont {McCallum}, \citenamefont {Sellars},\ and\ \citenamefont
  {Rogge}}]{deboo20}%
  \BibitemOpen
  \bibfield  {author} {\bibinfo {author} {\bibfnamefont {G.~G.}\ \bibnamefont
  {de~Boo}}, \bibinfo {author} {\bibfnamefont {C.}~\bibnamefont {Yin}},
  \bibinfo {author} {\bibfnamefont {M.}~\bibnamefont {Ran\v{c}i\'{c}}},
  \bibinfo {author} {\bibfnamefont {B.~C.}\ \bibnamefont {Johnson}}, \bibinfo
  {author} {\bibfnamefont {J.~C.}\ \bibnamefont {McCallum}}, \bibinfo {author}
  {\bibfnamefont {M.~J.}\ \bibnamefont {Sellars}},\ and\ \bibinfo {author}
  {\bibfnamefont {S.}~\bibnamefont {Rogge}},\ }\href
  {https://doi.org/10.1103/PhysRevB.102.155309} {\bibfield  {journal} {\bibinfo
   {journal} {Phys. Rev. B}\ }\textbf {\bibinfo {volume} {102}},\ \bibinfo
  {pages} {155309} (\bibinfo {year} {2020})}\BibitemShut {NoStop}%
\bibitem [{\citenamefont {Benton}\ \emph {et~al.}(1991)\citenamefont {Benton},
  \citenamefont {Michel}, \citenamefont {Kimerling}, \citenamefont {Jacobson},
  \citenamefont {Xie}, \citenamefont {Eaglesham}, \citenamefont {Fitzgerald},\
  and\ \citenamefont {Poate}}]{michel1991}%
  \BibitemOpen
  \bibfield  {author} {\bibinfo {author} {\bibfnamefont {J.~L.}\ \bibnamefont
  {Benton}}, \bibinfo {author} {\bibfnamefont {J.}~\bibnamefont {Michel}},
  \bibinfo {author} {\bibfnamefont {L.~C.}\ \bibnamefont {Kimerling}}, \bibinfo
  {author} {\bibfnamefont {D.~C.}\ \bibnamefont {Jacobson}}, \bibinfo {author}
  {\bibfnamefont {Y.~H.}\ \bibnamefont {Xie}}, \bibinfo {author} {\bibfnamefont
  {D.~J.}\ \bibnamefont {Eaglesham}}, \bibinfo {author} {\bibfnamefont {E.~A.}\
  \bibnamefont {Fitzgerald}},\ and\ \bibinfo {author} {\bibfnamefont {J.~M.}\
  \bibnamefont {Poate}},\ }\href {https://doi.org/10.1063/1.349381} {\bibfield
  {journal} {\bibinfo  {journal} {Journal of Applied Physics}\ }\textbf
  {\bibinfo {volume} {70}},\ \bibinfo {pages} {2667} (\bibinfo {year}
  {1991})}\BibitemShut {NoStop}%
\bibitem [{\citenamefont {Coffa}\ \emph {et~al.}(1993)\citenamefont {Coffa},
  \citenamefont {Priolo}, \citenamefont {Franzo}, \citenamefont {Bellani},
  \citenamefont {Carnera},\ and\ \citenamefont {Spinella}}]{coffa1993}%
  \BibitemOpen
  \bibfield  {author} {\bibinfo {author} {\bibfnamefont {S.}~\bibnamefont
  {Coffa}}, \bibinfo {author} {\bibfnamefont {F.}~\bibnamefont {Priolo}},
  \bibinfo {author} {\bibfnamefont {G.}~\bibnamefont {Franzo}}, \bibinfo
  {author} {\bibfnamefont {V.}~\bibnamefont {Bellani}}, \bibinfo {author}
  {\bibfnamefont {A.}~\bibnamefont {Carnera}},\ and\ \bibinfo {author}
  {\bibfnamefont {C.}~\bibnamefont {Spinella}},\ }\href
  {https://doi.org/10.1103/PhysRevB.48.11782} {\bibfield  {journal} {\bibinfo
  {journal} {Physical Review B}\ }\textbf {\bibinfo {volume} {48}},\ \bibinfo
  {pages} {11782} (\bibinfo {year} {1993})}\BibitemShut {NoStop}%
\bibitem [{\citenamefont {Carey}\ \emph {et~al.}(1996)\citenamefont {Carey},
  \citenamefont {Donegan}, \citenamefont {Barklie}, \citenamefont {Priolo},
  \citenamefont {Franzo},\ and\ \citenamefont {Coffa}}]{carey1996}%
  \BibitemOpen
  \bibfield  {author} {\bibinfo {author} {\bibfnamefont {J.~D.}\ \bibnamefont
  {Carey}}, \bibinfo {author} {\bibfnamefont {J.~F.}\ \bibnamefont {Donegan}},
  \bibinfo {author} {\bibfnamefont {R.~C.}\ \bibnamefont {Barklie}}, \bibinfo
  {author} {\bibfnamefont {F.}~\bibnamefont {Priolo}}, \bibinfo {author}
  {\bibfnamefont {G.}~\bibnamefont {Franzo}},\ and\ \bibinfo {author}
  {\bibfnamefont {S.}~\bibnamefont {Coffa}},\ }\href
  {https://doi.org/10.1063/1.117127} {\bibfield  {journal} {\bibinfo  {journal}
  {Applied Physics Letters}\ }\textbf {\bibinfo {volume} {69}},\ \bibinfo
  {pages} {3854} (\bibinfo {year} {1996})}\BibitemShut {NoStop}%
\bibitem [{\citenamefont {Wollman}\ \emph {et~al.}(2017)\citenamefont
  {Wollman}, \citenamefont {Verma}, \citenamefont {Beyer}, \citenamefont
  {Briggs}, \citenamefont {Korzh}, \citenamefont {Allmaras}, \citenamefont
  {Marsili}, \citenamefont {Lita}, \citenamefont {Mirin}, \citenamefont {Nam},\
  and\ \citenamefont {Shaw}}]{wollman2017}%
  \BibitemOpen
  \bibfield  {author} {\bibinfo {author} {\bibfnamefont {E.~E.}\ \bibnamefont
  {Wollman}}, \bibinfo {author} {\bibfnamefont {V.~B.}\ \bibnamefont {Verma}},
  \bibinfo {author} {\bibfnamefont {A.~D.}\ \bibnamefont {Beyer}}, \bibinfo
  {author} {\bibfnamefont {R.~M.}\ \bibnamefont {Briggs}}, \bibinfo {author}
  {\bibfnamefont {B.}~\bibnamefont {Korzh}}, \bibinfo {author} {\bibfnamefont
  {J.~P.}\ \bibnamefont {Allmaras}}, \bibinfo {author} {\bibfnamefont
  {F.}~\bibnamefont {Marsili}}, \bibinfo {author} {\bibfnamefont {A.~E.}\
  \bibnamefont {Lita}}, \bibinfo {author} {\bibfnamefont {R.~P.}\ \bibnamefont
  {Mirin}}, \bibinfo {author} {\bibfnamefont {S.~W.}\ \bibnamefont {Nam}},\
  and\ \bibinfo {author} {\bibfnamefont {M.~D.}\ \bibnamefont {Shaw}},\ }\href
  {https://doi.org/10.1364/Oe.25.026792} {\bibfield  {journal} {\bibinfo
  {journal} {Optics Express}\ }\textbf {\bibinfo {volume} {25}},\ \bibinfo
  {pages} {26792} (\bibinfo {year} {2017})}\BibitemShut {NoStop}%
\bibitem [{\citenamefont {Gol'tsman}\ \emph {et~al.}(2007)\citenamefont
  {Gol'tsman}, \citenamefont {Minaeva}, \citenamefont {Korneev}, \citenamefont
  {Tarkhov}, \citenamefont {Rubtsova}, \citenamefont {Divochiy}, \citenamefont
  {Milostnaya}, \citenamefont {Chulkova}, \citenamefont {Kaurova},
  \citenamefont {Voronov}, \citenamefont {Pan}, \citenamefont {Kitaygorsky},
  \citenamefont {Cross}, \citenamefont {Pearlman}, \citenamefont {Komissarov},
  \citenamefont {Slysz}, \citenamefont {Wegrzecki}, \citenamefont {Grabiec},\
  and\ \citenamefont {Sobolewski}}]{goltsman2007}%
  \BibitemOpen
  \bibfield  {author} {\bibinfo {author} {\bibfnamefont {G.}~\bibnamefont
  {Gol'tsman}}, \bibinfo {author} {\bibfnamefont {O.}~\bibnamefont {Minaeva}},
  \bibinfo {author} {\bibfnamefont {A.}~\bibnamefont {Korneev}}, \bibinfo
  {author} {\bibfnamefont {M.}~\bibnamefont {Tarkhov}}, \bibinfo {author}
  {\bibfnamefont {I.}~\bibnamefont {Rubtsova}}, \bibinfo {author}
  {\bibfnamefont {A.}~\bibnamefont {Divochiy}}, \bibinfo {author}
  {\bibfnamefont {I.}~\bibnamefont {Milostnaya}}, \bibinfo {author}
  {\bibfnamefont {G.}~\bibnamefont {Chulkova}}, \bibinfo {author}
  {\bibfnamefont {N.}~\bibnamefont {Kaurova}}, \bibinfo {author} {\bibfnamefont
  {B.}~\bibnamefont {Voronov}}, \bibinfo {author} {\bibfnamefont
  {D.}~\bibnamefont {Pan}}, \bibinfo {author} {\bibfnamefont {J.}~\bibnamefont
  {Kitaygorsky}}, \bibinfo {author} {\bibfnamefont {A.}~\bibnamefont {Cross}},
  \bibinfo {author} {\bibfnamefont {A.}~\bibnamefont {Pearlman}}, \bibinfo
  {author} {\bibfnamefont {I.}~\bibnamefont {Komissarov}}, \bibinfo {author}
  {\bibfnamefont {W.}~\bibnamefont {Slysz}}, \bibinfo {author} {\bibfnamefont
  {M.}~\bibnamefont {Wegrzecki}}, \bibinfo {author} {\bibfnamefont
  {P.}~\bibnamefont {Grabiec}},\ and\ \bibinfo {author} {\bibfnamefont
  {R.}~\bibnamefont {Sobolewski}},\ }\href
  {https://doi.org/10.1109/Tasc.2007.898252} {\bibfield  {journal} {\bibinfo
  {journal} {Ieee Transactions on Applied Superconductivity}\ }\textbf
  {\bibinfo {volume} {17}},\ \bibinfo {pages} {246} (\bibinfo {year}
  {2007})}\BibitemShut {NoStop}%
\bibitem [{\citenamefont {Marsili}\ \emph {et~al.}(2012)\citenamefont
  {Marsili}, \citenamefont {Bellei}, \citenamefont {Najafi}, \citenamefont
  {Dane}, \citenamefont {Dauler}, \citenamefont {Molnar},\ and\ \citenamefont
  {Berggren}}]{marsili2012}%
  \BibitemOpen
  \bibfield  {author} {\bibinfo {author} {\bibfnamefont {F.}~\bibnamefont
  {Marsili}}, \bibinfo {author} {\bibfnamefont {F.}~\bibnamefont {Bellei}},
  \bibinfo {author} {\bibfnamefont {F.}~\bibnamefont {Najafi}}, \bibinfo
  {author} {\bibfnamefont {A.~E.}\ \bibnamefont {Dane}}, \bibinfo {author}
  {\bibfnamefont {E.~A.}\ \bibnamefont {Dauler}}, \bibinfo {author}
  {\bibfnamefont {R.~J.}\ \bibnamefont {Molnar}},\ and\ \bibinfo {author}
  {\bibfnamefont {K.~K.}\ \bibnamefont {Berggren}},\ }\href
  {https://doi.org/10.1021/nl302245n} {\bibfield  {journal} {\bibinfo
  {journal} {Nano Letters}\ }\textbf {\bibinfo {volume} {12}},\ \bibinfo
  {pages} {4799} (\bibinfo {year} {2012})}\BibitemShut {NoStop}%
\bibitem [{\citenamefont {Chen}\ \emph {et~al.}(2017)\citenamefont {Chen},
  \citenamefont {Schwarzer}, \citenamefont {Verma}, \citenamefont {Stevens},
  \citenamefont {Marsili}, \citenamefont {Mirin}, \citenamefont {Nam},\ and\
  \citenamefont {Wodtke}}]{chen2017}%
  \BibitemOpen
  \bibfield  {author} {\bibinfo {author} {\bibfnamefont {L.}~\bibnamefont
  {Chen}}, \bibinfo {author} {\bibfnamefont {D.}~\bibnamefont {Schwarzer}},
  \bibinfo {author} {\bibfnamefont {V.~B.}\ \bibnamefont {Verma}}, \bibinfo
  {author} {\bibfnamefont {M.~J.}\ \bibnamefont {Stevens}}, \bibinfo {author}
  {\bibfnamefont {F.}~\bibnamefont {Marsili}}, \bibinfo {author} {\bibfnamefont
  {R.~P.}\ \bibnamefont {Mirin}}, \bibinfo {author} {\bibfnamefont {S.~W.}\
  \bibnamefont {Nam}},\ and\ \bibinfo {author} {\bibfnamefont {A.~M.}\
  \bibnamefont {Wodtke}},\ }\href
  {https://doi.org/10.1021/acs.accounts.7b00071} {\bibfield  {journal}
  {\bibinfo  {journal} {Accounts of Chemical Research}\ }\textbf {\bibinfo
  {volume} {50}},\ \bibinfo {pages} {1400} (\bibinfo {year}
  {2017})}\BibitemShut {NoStop}%
\bibitem [{\citenamefont {Taguchi}\ \emph {et~al.}(1998)\citenamefont
  {Taguchi}, \citenamefont {Takahei}, \citenamefont {Matsuoka},\ and\
  \citenamefont {Tohno}}]{Taguchi1998}%
  \BibitemOpen
  \bibfield  {author} {\bibinfo {author} {\bibfnamefont {A.}~\bibnamefont
  {Taguchi}}, \bibinfo {author} {\bibfnamefont {K.}~\bibnamefont {Takahei}},
  \bibinfo {author} {\bibfnamefont {M.}~\bibnamefont {Matsuoka}},\ and\
  \bibinfo {author} {\bibfnamefont {S.}~\bibnamefont {Tohno}},\ }\href
  {https://doi.org/10.1063/1.368673} {\bibfield  {journal} {\bibinfo  {journal}
  {Journal of Applied Physics}\ }\textbf {\bibinfo {volume} {84}},\ \bibinfo
  {pages} {4471} (\bibinfo {year} {1998})}\BibitemShut {NoStop}%
\bibitem [{\citenamefont {Priolo}\ \emph {et~al.}(1998)\citenamefont {Priolo},
  \citenamefont {Franzo}, \citenamefont {Coffa},\ and\ \citenamefont
  {Carnera}}]{Priolo1998}%
  \BibitemOpen
  \bibfield  {author} {\bibinfo {author} {\bibfnamefont {F.}~\bibnamefont
  {Priolo}}, \bibinfo {author} {\bibfnamefont {G.}~\bibnamefont {Franzo}},
  \bibinfo {author} {\bibfnamefont {S.}~\bibnamefont {Coffa}},\ and\ \bibinfo
  {author} {\bibfnamefont {A.}~\bibnamefont {Carnera}},\ }\href
  {https://doi.org/10.1103/PhysRevB.57.4443} {\bibfield  {journal} {\bibinfo
  {journal} {Physical Review B}\ }\textbf {\bibinfo {volume} {57}},\ \bibinfo
  {pages} {4443} (\bibinfo {year} {1998})}\BibitemShut {NoStop}%
\bibitem [{\citenamefont {Dibos}\ \emph {et~al.}(2018)\citenamefont {Dibos},
  \citenamefont {Raha}, \citenamefont {Phenicie},\ and\ \citenamefont
  {Thompson}}]{dibos2018}%
  \BibitemOpen
  \bibfield  {author} {\bibinfo {author} {\bibfnamefont {M.}~\bibnamefont
  {Dibos}}, \bibinfo {author} {\bibfnamefont {M.}~\bibnamefont {Raha}},
  \bibinfo {author} {\bibfnamefont {C.~M.}\ \bibnamefont {Phenicie}},\ and\
  \bibinfo {author} {\bibfnamefont {J.~D.}\ \bibnamefont {Thompson}},\
  }\bibfield  {journal} {\bibinfo  {journal} {Physical Review Letters}\
  }\textbf {\bibinfo {volume} {120}},\ \href
  {https://doi.org/10.1103/PhysRevLett.120.243601}
  {10.1103/PhysRevLett.120.243601} (\bibinfo {year} {2018})\BibitemShut
  {NoStop}%
\bibitem [{\citenamefont {B\"{o}ttger}\ \emph {et~al.}(2006)\citenamefont
  {B\"{o}ttger}, \citenamefont {Sun}, \citenamefont {Thiel},\ and\
  \citenamefont {Cone}}]{bottger2006}%
  \BibitemOpen
  \bibfield  {author} {\bibinfo {author} {\bibfnamefont {T.}~\bibnamefont
  {B\"{o}ttger}}, \bibinfo {author} {\bibfnamefont {Y.}~\bibnamefont {Sun}},
  \bibinfo {author} {\bibfnamefont {C.~W.}\ \bibnamefont {Thiel}},\ and\
  \bibinfo {author} {\bibfnamefont {R.~L.}\ \bibnamefont {Cone}},\ }\bibfield
  {journal} {\bibinfo  {journal} {Physical Review B}\ }\textbf {\bibinfo
  {volume} {74}},\ \href {https://doi.org/ARTN 075107
  10.1103/PhysRevB.74.075107} {ARTN 075107 10.1103/PhysRevB.74.075107}
  (\bibinfo {year} {2006})\BibitemShut {NoStop}%
\bibitem [{\citenamefont {Vinh}\ \emph {et~al.}(2004)\citenamefont {Vinh},
  \citenamefont {Przybyli\ifmmode~\acute{n}\else \'{n}\fi{}ska}, \citenamefont
  {Krasil'nik},\ and\ \citenamefont {Gregorkiewicz}}]{vinh04}%
  \BibitemOpen
  \bibfield  {author} {\bibinfo {author} {\bibfnamefont {N.~Q.}\ \bibnamefont
  {Vinh}}, \bibinfo {author} {\bibfnamefont {H.}~\bibnamefont
  {Przybyli\ifmmode~\acute{n}\else \'{n}\fi{}ska}}, \bibinfo {author}
  {\bibfnamefont {Z.~F.}\ \bibnamefont {Krasil'nik}},\ and\ \bibinfo {author}
  {\bibfnamefont {T.}~\bibnamefont {Gregorkiewicz}},\ }\bibfield  {journal}
  {\bibinfo  {journal} {Phys. Rev. B}\ }\textbf {\bibinfo {volume} {70}},\
  \href {https://doi.org/10.1103/PhysRevB.70.115332}
  {10.1103/PhysRevB.70.115332} (\bibinfo {year} {2004})\BibitemShut {NoStop}%
\bibitem [{\citenamefont {Szabo}(1975)}]{Szabo1975}%
  \BibitemOpen
  \bibfield  {author} {\bibinfo {author} {\bibfnamefont {A.}~\bibnamefont
  {Szabo}},\ }\href {https://doi.org/10.1103/PhysRevB.11.4512} {\bibfield
  {journal} {\bibinfo  {journal} {Physical Review B}\ }\textbf {\bibinfo
  {volume} {11}},\ \bibinfo {pages} {4512} (\bibinfo {year}
  {1975})}\BibitemShut {NoStop}%
\bibitem [{\citenamefont {Moerner}(1988)}]{moerner1988}%
  \BibitemOpen
  \bibfield  {author} {\bibinfo {author} {\bibfnamefont {W.~E.}\ \bibnamefont
  {Moerner}},\ }\href {https://doi.org/10.1007/978-3-642-83290-1} {\emph
  {\bibinfo {title} {Persistent Spectral Hole-Burning: Science and
  Applications}}}\ (\bibinfo  {publisher} {Springer},\ \bibinfo {year} {1988})\
  p.~\bibinfo {pages} {5}\BibitemShut {NoStop}%
\bibitem [{\citenamefont {Huang}\ \emph {et~al.}(2001)\citenamefont {Huang},
  \citenamefont {Mortier},\ and\ \citenamefont {Auzel}}]{Huang2001}%
  \BibitemOpen
  \bibfield  {author} {\bibinfo {author} {\bibfnamefont {Y.~D.}\ \bibnamefont
  {Huang}}, \bibinfo {author} {\bibfnamefont {M.}~\bibnamefont {Mortier}},\
  and\ \bibinfo {author} {\bibfnamefont {F.}~\bibnamefont {Auzel}},\ }\href
  {https://doi.org/10.1016/S0925-3467(00)00039-2} {\bibfield  {journal}
  {\bibinfo  {journal} {Optical Materials}\ }\textbf {\bibinfo {volume} {15}},\
  \bibinfo {pages} {243} (\bibinfo {year} {2001})}\BibitemShut {NoStop}%
\bibitem [{\citenamefont {Coffa}\ \emph {et~al.}(1994)\citenamefont {Coffa},
  \citenamefont {Franzo}, \citenamefont {Priolo}, \citenamefont {Polman},\ and\
  \citenamefont {Serna}}]{Coffa1994}%
  \BibitemOpen
  \bibfield  {author} {\bibinfo {author} {\bibfnamefont {S.}~\bibnamefont
  {Coffa}}, \bibinfo {author} {\bibfnamefont {G.}~\bibnamefont {Franzo}},
  \bibinfo {author} {\bibfnamefont {F.}~\bibnamefont {Priolo}}, \bibinfo
  {author} {\bibfnamefont {A.}~\bibnamefont {Polman}},\ and\ \bibinfo {author}
  {\bibfnamefont {R.}~\bibnamefont {Serna}},\ }\href
  {https://doi.org/10.1103/PhysRevB.49.16313} {\bibfield  {journal} {\bibinfo
  {journal} {Physical Review B}\ }\textbf {\bibinfo {volume} {49}},\ \bibinfo
  {pages} {16313} (\bibinfo {year} {1994})}\BibitemShut {NoStop}%
\bibitem [{\citenamefont {Priolo}\ \emph {et~al.}(1995)\citenamefont {Priolo},
  \citenamefont {Franzo}, \citenamefont {Coffa}, \citenamefont {Polman},
  \citenamefont {Libertino}, \citenamefont {Barklie},\ and\ \citenamefont
  {Carey}}]{Priolo1995}%
  \BibitemOpen
  \bibfield  {author} {\bibinfo {author} {\bibfnamefont {F.}~\bibnamefont
  {Priolo}}, \bibinfo {author} {\bibfnamefont {G.}~\bibnamefont {Franzo}},
  \bibinfo {author} {\bibfnamefont {S.}~\bibnamefont {Coffa}}, \bibinfo
  {author} {\bibfnamefont {A.}~\bibnamefont {Polman}}, \bibinfo {author}
  {\bibfnamefont {S.}~\bibnamefont {Libertino}}, \bibinfo {author}
  {\bibfnamefont {R.}~\bibnamefont {Barklie}},\ and\ \bibinfo {author}
  {\bibfnamefont {D.}~\bibnamefont {Carey}},\ }\href
  {https://doi.org/10.1063/1.359904} {\bibfield  {journal} {\bibinfo  {journal}
  {Journal of Applied Physics}\ }\textbf {\bibinfo {volume} {78}},\ \bibinfo
  {pages} {3874} (\bibinfo {year} {1995})}\BibitemShut {NoStop}%
\bibitem [{\citenamefont {Wu}\ \emph {et~al.}(1997)\citenamefont {Wu},
  \citenamefont {White}, \citenamefont {Hommerich}, \citenamefont {Namavar},\
  and\ \citenamefont {CreminsCosta}}]{Wu1997}%
  \BibitemOpen
  \bibfield  {author} {\bibinfo {author} {\bibfnamefont {X.}~\bibnamefont
  {Wu}}, \bibinfo {author} {\bibfnamefont {R.}~\bibnamefont {White}}, \bibinfo
  {author} {\bibfnamefont {U.}~\bibnamefont {Hommerich}}, \bibinfo {author}
  {\bibfnamefont {F.}~\bibnamefont {Namavar}},\ and\ \bibinfo {author}
  {\bibfnamefont {A.~M.}\ \bibnamefont {CreminsCosta}},\ }\href
  {https://doi.org/10.1016/S0022-2313(96)00098-1} {\bibfield  {journal}
  {\bibinfo  {journal} {Journal of Luminescence}\ }\textbf {\bibinfo {volume}
  {71}},\ \bibinfo {pages} {13} (\bibinfo {year} {1997})}\BibitemShut {NoStop}%
\bibitem [{\citenamefont {Vinh}\ \emph {et~al.}(2005)\citenamefont {Vinh},
  \citenamefont {Minissale}, \citenamefont {Andreev},\ and\ \citenamefont
  {Gregorkiewicz}}]{Vinh2005}%
  \BibitemOpen
  \bibfield  {author} {\bibinfo {author} {\bibfnamefont {N.~Q.}\ \bibnamefont
  {Vinh}}, \bibinfo {author} {\bibfnamefont {S.}~\bibnamefont {Minissale}},
  \bibinfo {author} {\bibfnamefont {B.~A.}\ \bibnamefont {Andreev}},\ and\
  \bibinfo {author} {\bibfnamefont {T.}~\bibnamefont {Gregorkiewicz}},\ }\href
  {https://doi.org/10.1088/0953-8984/17/22/006} {\bibfield  {journal} {\bibinfo
   {journal} {Journal of Physics-Condensed Matter}\ }\textbf {\bibinfo {volume}
  {17}},\ \bibinfo {pages} {S2191} (\bibinfo {year} {2005})}\BibitemShut
  {NoStop}%
\bibitem [{\citenamefont {Dodson}\ and\ \citenamefont {Zia}(2012)}]{dodson12}%
  \BibitemOpen
  \bibfield  {author} {\bibinfo {author} {\bibfnamefont {C.~M.}\ \bibnamefont
  {Dodson}}\ and\ \bibinfo {author} {\bibfnamefont {R.}~\bibnamefont {Zia}},\
  }\href {https://doi.org/10.1103/PhysRevB.86.125102} {\bibfield  {journal}
  {\bibinfo  {journal} {Phys. Rev. B}\ }\textbf {\bibinfo {volume} {86}},\
  \bibinfo {pages} {125102} (\bibinfo {year} {2012})}\BibitemShut {NoStop}%
\bibitem [{\citenamefont {Yamaguchi}\ \emph {et~al.}(2008)\citenamefont
  {Yamaguchi}, \citenamefont {Asano}, \citenamefont {Fujita},\ and\
  \citenamefont {Noda}}]{yamaguchi2008}%
  \BibitemOpen
  \bibfield  {author} {\bibinfo {author} {\bibfnamefont {M.}~\bibnamefont
  {Yamaguchi}}, \bibinfo {author} {\bibfnamefont {T.}~\bibnamefont {Asano}},
  \bibinfo {author} {\bibfnamefont {M.}~\bibnamefont {Fujita}},\ and\ \bibinfo
  {author} {\bibfnamefont {S.}~\bibnamefont {Noda}},\ }\href
  {https://doi.org/10.1002/pssc.200779247} {\bibfield  {journal} {\bibinfo
  {journal} {Physica Status Solidi C - Current Topics in Solid State Physics,
  Vol 5, No 9}\ }\textbf {\bibinfo {volume} {5}},\ \bibinfo {pages} {2828}
  (\bibinfo {year} {2008})}\BibitemShut {NoStop}%
\bibitem [{\citenamefont {Zhou}\ \emph {et~al.}(2019)\citenamefont {Zhou},
  \citenamefont {Zheng}, \citenamefont {Fang}, \citenamefont {Xu},\ and\
  \citenamefont {Majumdar}}]{zhou2019}%
  \BibitemOpen
  \bibfield  {author} {\bibinfo {author} {\bibfnamefont {J.}~\bibnamefont
  {Zhou}}, \bibinfo {author} {\bibfnamefont {J.~J.}\ \bibnamefont {Zheng}},
  \bibinfo {author} {\bibfnamefont {Z.~R.}\ \bibnamefont {Fang}}, \bibinfo
  {author} {\bibfnamefont {P.~P.}\ \bibnamefont {Xu}},\ and\ \bibinfo {author}
  {\bibfnamefont {A.}~\bibnamefont {Majumdar}},\ }\href
  {https://doi.org/10.1364/Oe.27.030692} {\bibfield  {journal} {\bibinfo
  {journal} {Optics Express}\ }\textbf {\bibinfo {volume} {27}},\ \bibinfo
  {pages} {30692} (\bibinfo {year} {2019})}\BibitemShut {NoStop}%
\bibitem [{\citenamefont {Miura}\ \emph {et~al.}(2014)\citenamefont {Miura},
  \citenamefont {Imamura}, \citenamefont {Ohta}, \citenamefont {Ishii},
  \citenamefont {Liu}, \citenamefont {Shimada}, \citenamefont {Iwamoto},
  \citenamefont {Arakawa},\ and\ \citenamefont {Kato}}]{miura2014}%
  \BibitemOpen
  \bibfield  {author} {\bibinfo {author} {\bibfnamefont {R.}~\bibnamefont
  {Miura}}, \bibinfo {author} {\bibfnamefont {S.}~\bibnamefont {Imamura}},
  \bibinfo {author} {\bibfnamefont {R.}~\bibnamefont {Ohta}}, \bibinfo {author}
  {\bibfnamefont {A.}~\bibnamefont {Ishii}}, \bibinfo {author} {\bibfnamefont
  {X.}~\bibnamefont {Liu}}, \bibinfo {author} {\bibfnamefont {T.}~\bibnamefont
  {Shimada}}, \bibinfo {author} {\bibfnamefont {S.}~\bibnamefont {Iwamoto}},
  \bibinfo {author} {\bibfnamefont {Y.}~\bibnamefont {Arakawa}},\ and\ \bibinfo
  {author} {\bibfnamefont {Y.~K.}\ \bibnamefont {Kato}},\ }\bibfield  {journal}
  {\bibinfo  {journal} {Nature Communications}\ }\textbf {\bibinfo {volume}
  {5}},\ \href {https://doi.org/ARTN 5580 10.1038/ncomms6580} {ARTN 5580
  10.1038/ncomms6580} (\bibinfo {year} {2014})\BibitemShut {NoStop}%
\bibitem [{\citenamefont {Marsili}\ \emph {et~al.}(2013)\citenamefont
  {Marsili}, \citenamefont {Verma}, \citenamefont {Stern}, \citenamefont
  {Harrington}, \citenamefont {Lita}, \citenamefont {Gerrits}, \citenamefont
  {Vayshenker}, \citenamefont {Baek}, \citenamefont {Shaw}, \citenamefont
  {Mirin},\ and\ \citenamefont {Nam}}]{marsili2013}%
  \BibitemOpen
  \bibfield  {author} {\bibinfo {author} {\bibfnamefont {F.}~\bibnamefont
  {Marsili}}, \bibinfo {author} {\bibfnamefont {V.~B.}\ \bibnamefont {Verma}},
  \bibinfo {author} {\bibfnamefont {J.~A.}\ \bibnamefont {Stern}}, \bibinfo
  {author} {\bibfnamefont {S.}~\bibnamefont {Harrington}}, \bibinfo {author}
  {\bibfnamefont {A.~E.}\ \bibnamefont {Lita}}, \bibinfo {author}
  {\bibfnamefont {T.}~\bibnamefont {Gerrits}}, \bibinfo {author} {\bibfnamefont
  {I.}~\bibnamefont {Vayshenker}}, \bibinfo {author} {\bibfnamefont
  {B.}~\bibnamefont {Baek}}, \bibinfo {author} {\bibfnamefont {M.~D.}\
  \bibnamefont {Shaw}}, \bibinfo {author} {\bibfnamefont {R.~P.}\ \bibnamefont
  {Mirin}},\ and\ \bibinfo {author} {\bibfnamefont {S.~W.}\ \bibnamefont
  {Nam}},\ }\href {https://doi.org/10.1038/Nphoton.2013.13} {\bibfield
  {journal} {\bibinfo  {journal} {Nature Photonics}\ }\textbf {\bibinfo
  {volume} {7}},\ \bibinfo {pages} {210} (\bibinfo {year} {2013})}\BibitemShut
  {NoStop}%
\end{thebibliography}%

%%%%%%%%%%%%%%%%%%%%%%%%%%%%%%%%%%%%%%%%%%%%%%%%%%%%%%%%%%%%%%%%%%%%%%%%%%%%%%%

\clearpage
\beginsupplement
\onecolumngrid
\begin{center}
%\textbf{\large Supplemental Material: Photoluminescence excitation spectroscopy of Er in Si via in-situ single photon detection}
\section*{Supplemental Material: Photoluminescence excitation spectroscopy of Er in Si via in-situ single photon detection}
\end{center}
\twocolumngrid

\subsection{SSPD}
The SSPD has been fabricated at the ANFF NSW-node using the technique presented in Ref. \cite{marsili2013}. The system detection efficiency of the SSPD is obtained by calibrating the ratio between the fiber to the SSPD and the power meter. The power meter will give the theoretical value of the amount of photons $N_{photons}$ expected on the SSPD. The system detection efficiency is then calculated using \SI{100}{\percent}$\cdot(\text{CR}-\text{DCR})/N_{photons}$, where CR is the count rate when the laser is on and DCR is the dark count rate when the laser is turned off. At the bias current used for this experiment, the SSPD has a system detection efficiency of \SI{66.27}{\percent} at \SI{1550}{\nano\metre}.

\subsection{Photoluminescent excitation spectrum}
\begin{figure}
	\centering
        \includegraphics[width=1.0\columnwidth]{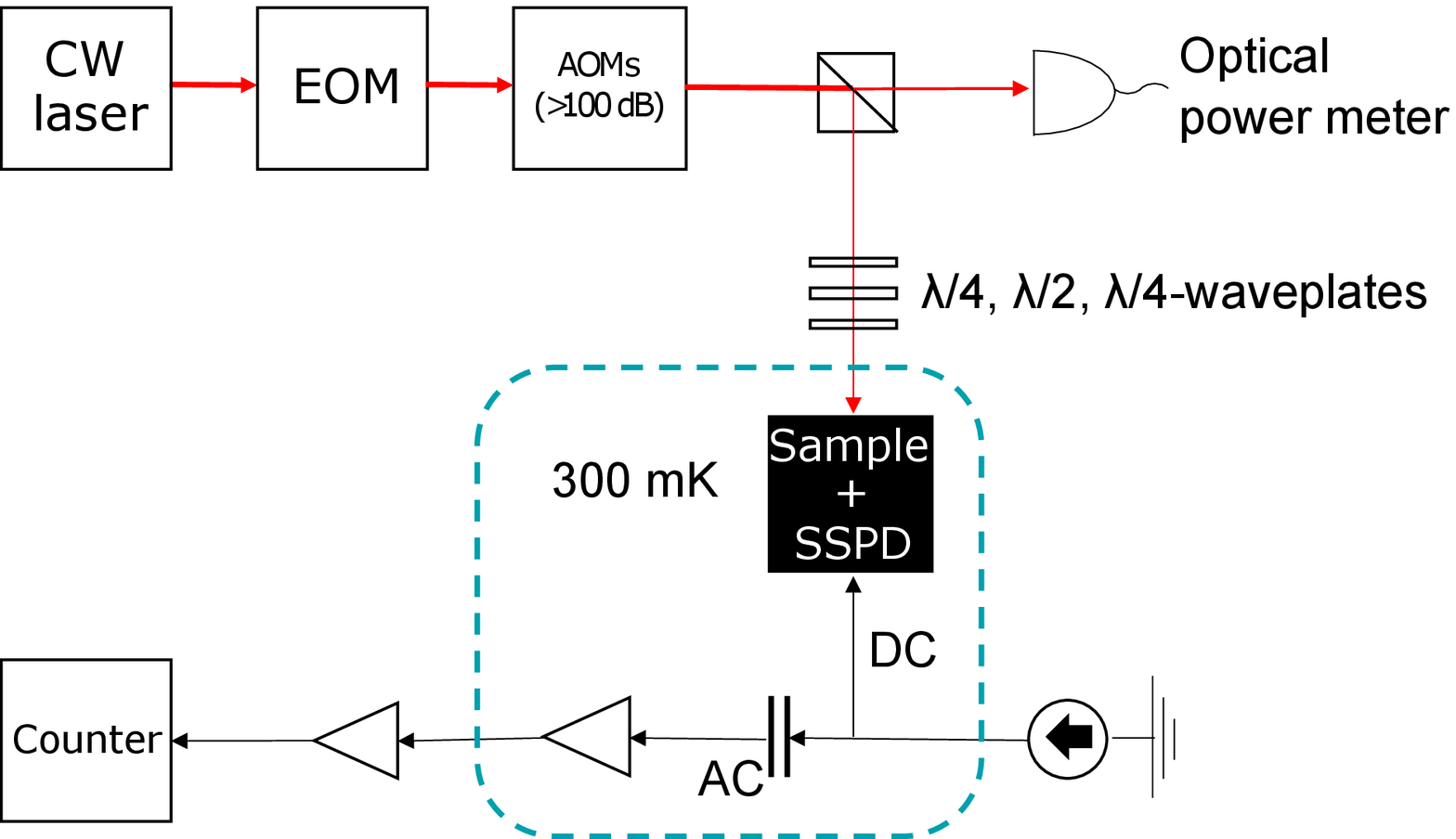}
        \caption{Experimental setup}
        \label{fig:setupa}
\end{figure}

\begin{figure}
    \centering
        \includegraphics[width=1.0\columnwidth]{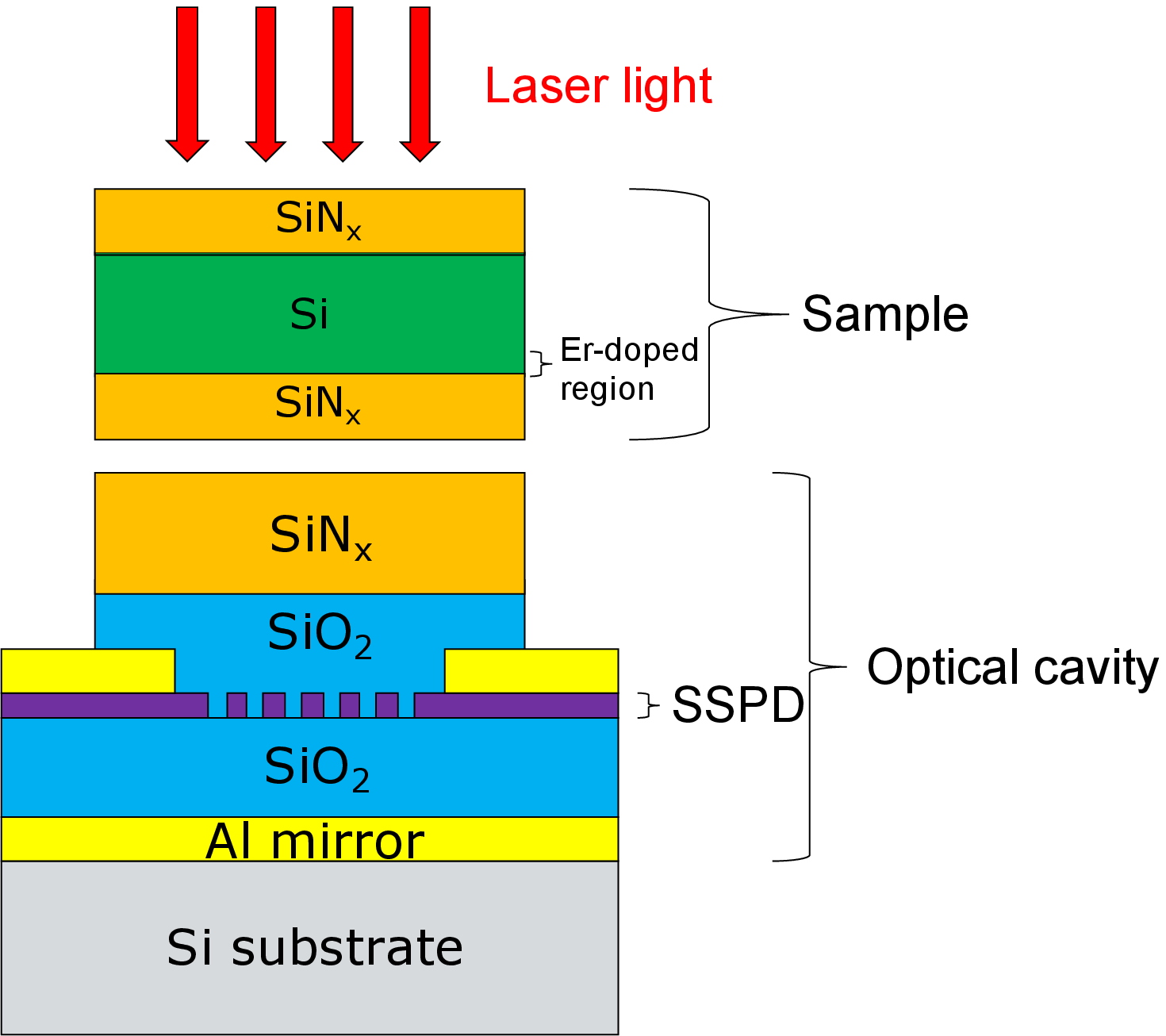}
        \caption{Sample and SSPD}
        \label{fig:setupb}
\end{figure}

A schematic of the setup is shown in Fig. \ref{fig:setupa} and the in situ sample and \ac{SSPD} in Fig. \ref{fig:setupb}. The CW laser is the Pure Photonics PPCL550 which has been attenuated by a Thorlabs fixed attenuator (FA15T) and Thorlabs polarization maintaining variable optical attenuator (VOA50PM-APC). Afterwards, the light is split using a Thorlabs 90/10 SM beamsplitter where the low output port is connected to a Bristol 671B wavelength meter. Following the beamsplitter is a Brimrose AOM (AMM-100-10-50-1550-2FP-SM) and an AA optoelectronic AOM (MT80-IIR30-Fio-SM0) connected to their corresponding drivers which is pulsed using the Keysight 33520B with a delay between the two channels to account for the optical fibre length between the two AOMs. A Thorlabs 90/10 beamsplitter was used to direct \SI{10}{\percent} of the light to a Thorlabs PM100D power meter with an Thorlabs S154C photodiode. The higher output end is directed to polarization paddles (Thorlabs FPC031) to adjust the polarization for the magnetic field measurements. 

The fiber is coupled to an Oxford HelioxVL which resides in an Oxford magnet dewar. The heliox contains the sample, SSPD, bias-tee and a homemade HEMT amplifier. The bias-tee ensures that a bias current can be applied to the SSPD and a \SI{2}{\nano\second} pulse can be read out via coaxial lines. The SSPD AC signal is amplified at room temperature using two Mini-Circuits ZFL-1000LN+. The pulse was stretched using a homemade double-comparator pulse stretcher and read out using a National Instruments 6602 counter card. The counter card was initialised to measure the number of counts in \SI{10}{\micro\second} bins.


\subsection{Spectral hole burning setup}\label{sec:shb}
To perform the spectral hole burning measurements, polarization paddles (Thorlabs FPC032) and Covega EOM were connected in series right before the AOMs. A voltage source was connected to the DC port, a Stanford Research Systems SG384 RF source and a Keysight N512B vector source SMA outputs are combined using an Mini-Circuits RF combiner and amplified before the RF signal reaches the RF port of the EOM. The SRS SG384 is set to \SI{2.5}{\giga\hertz} and \SI{0}{\decibel\meter} which is pulse triggered by an external pulse. The Keysight N512B is set to (\num{2.5}+$\Delta f$)\SI{}{\giga\hertz} where $\Delta f$ is the detuning frequency, additionally the vector source is triggered on its internal clock, which is triggered externally. To pulse both RF sources externally, a BNC T-piece was connected to the sync output of the Keysight 33520B used to pulse the AOMs. One end was directly coupled to the external pulse input on the Keysight N512B and an internal delay is set to output the RF pulse after the trailing edge of the SRS SG384 pulse. The other end of the T-piece is connected to the external input of a Keysight 33511B, which outputs an inverted pulse signal and its channel output is connected to the external input of the SRS SG384. To align the pulses, delays were implemented in the system which has been calibrated by coupling the fiber to the cryogenic system to a \SI{125}{\mega\hertz} \ce{InGaAs} photodiode, connected to Tektronix MS054 oscilloscope. The RF sources were pulsed in such a way that the SRS SG384 was continuously on except when the Keysight N512B outputs a pulse for \SI{150}{\micro\second}, where the last \SI{50}{\micro\second} continue after the AOM extinguished the light.

The EOM was initialised by turning the RF sources off and the laser on and adjusting the DC voltage until it shows maximum extinction of the carrier. Additionally, the polarization was adjusted to extinguish the carrier further. The RF signal of the SRS SG384 is turned on to create the sidebands and the polarization is finetuned until the ratio of the optical power when the RF source is on and off is maximized. The extinction of the carrier when the RF source is on is confirmed by sweeping both sidebands over an inhomogeneous peak. The extinction ratio is over \SI{15}{\decibel} in our spectral hole burning measurements.

The ratio between the two ends was calibrated using the PM100D power meter to calculate the amount of light on the sample. Using the calibrated detection efficiency of the SSPD, the transmission through the sample could be calculated which is estimated to be \SI{96}{\percent}. Dividing the value by \num{2} gives the approximated power of one sideband on the ions.

\subsection{Lifetime measurements}

The lifetimes are obtained by subtracting the offresonant time trace from a nearby onresonant time trace. A list of the chosen offresonant wavelengths is given in \ref{tab:offresresonances}.

\renewcommand\arraystretch{0.5}
\begin{longtable}{p{0.5\columnwidth} p{0.5\columnwidth}}
    \caption{Overview table}\label{tab:offresresonances}\\
        \toprule
         Resonant wavelength (\si{\nano\metre}) & Offresonant wavelength (\si{\nano\meter})\\
         \midrule
         \num{1539.948} &\num{1539.900}\\
         \num{1538.685} &\num{1538.635}\\
         \num{1538.242} &\num{1538.195} \\
         \num{1537.847} &\num{1537.803}\\
         \num{1537.651} &\num{1537.625}\\
         \num{1537.219} &\num{1537.182}\\
         \num{1536.762} &\num{1536.754}\\
         \num{1536.708} &\num{1536.675}\\
         \num{1536.683} &\num{1536.672}\\
         \num{1536.517} &\num{1536.553}\\
         \num{1536.489} &\num{1536.453}\\
         \num{1536.215} &\num{1536.174}\\
         \num{1536.139} &\num{1536.176}\\
         \num{1535.899} &\num{1535.861}\\
         \num{1535.199} &\num{1535.183}\\
%         \num{1535.007} &\num{1.827} & \acs{SNR} too low  &\\
         \num{1534.925} &\num{1534.959}\\
         \num{1534.796} &\num{1534.774}\\
         \num{1534.673} &\num{1534.610}\\
         \num{1534.507} &\num{1534.530}\\
         \num{1534.469} &\num{1534.435}\\
         \num{1534.373} &\num{1534.315}\\
         \num{1534.194} &\num{1534.125}\\
         \num{1534.079} &\num{1534.124}\\
         \num{1533.984} &\num{1533.952}\\
         \num{1533.884} &\num{1533.856}\\
         \num{1533.090} &\num{1533.116}\\
%         \num{1533.069} &\num{0.757} & \acs{SNR} too low  &\\
         \num{1532.793} &\num{1532.770}\\
%         \num{1532.371} &\num{2.146} &\num{0.015}  &\\
         \num{1532.254} &\num{1532.177}\\
         \num{1531.885} &\num{1531.850}\\
         \num{1530.063} &\num{1530.096}\\
         \num{1530.039} &\num{1529.996}\\
         \num{1529.958} &\num{1529.934}\\
         \num{1529.917} &\num{1529.900}\\
         \num{1529.656} &\num{1529.599}\\
         \num{1528.384} &\num{1528.292}\\
         \num{1527.960} &\num{1527.900}\\
         \num{1527.856} &\num{1527.904}\\
         \num{1527.738} &\num{1527.675}\\
         \num{1527.564} &\num{1527.461}\\
         \num{1526.774} &\num{1526.703}\\
         \num{1526.572} &\num{1526.521}\\
         \num{1526.170} &\num{1526.208}\\
         \num{1526.088} &\num{1526.129}\\
         \num{1525.885} &\num{1515.932}\\
         \num{1525.848} &\num{1525.817}\\
         \num{1525.750} &\num{1525.716}\\
         \num{1525.679} &\num{1525.663}\\
         \num{1525.514} &\num{1525.445}\\
         \num{1524.577} &\num{1524.542}\\
         \num{1524.360} &\num{1524.315}\\
         \num{1523.753} &\num{1523.713}\\
         \num{1523.532} &\num{1523.466}\\
         \num{1523.126} &\num{1523.073}\\
         \num{1523.052} &\num{1523.077}\\
         \num{1522.918} &\num{1522.891}\\
         \num{1522.834} &\num{1522.856}\\
         \num{1522.797} &\num{1522.767}\\
         \num{1522.400} &\num{1522.382}\\
         \num{1522.294} &\num{1522.244}\\
         \num{1522.113} &\num{1522.122}\\
         \num{1522.086} &\num{1522.067}\\
         \num{1522.025} &\num{1522.045}\\
         \num{1521.994} &\num{1521.971}\\
         \num{1521.817} &\num{1521.766}\\
         \num{1521.409} &\num{1521.447}\\
         \num{1520.926} &\num{1521.002}\\
         \num{1520.408} &\num{1520.490}\\
         \num{1520.097} &\num{1520.490}\\
         \num{1519.788} &\num{1519.698}\\
         \num{1518.042} &\num{1517.943}\\
        \bottomrule
\end{longtable}

\begin{figure}[b]
	\centering
    \includegraphics[width=1.0\columnwidth]{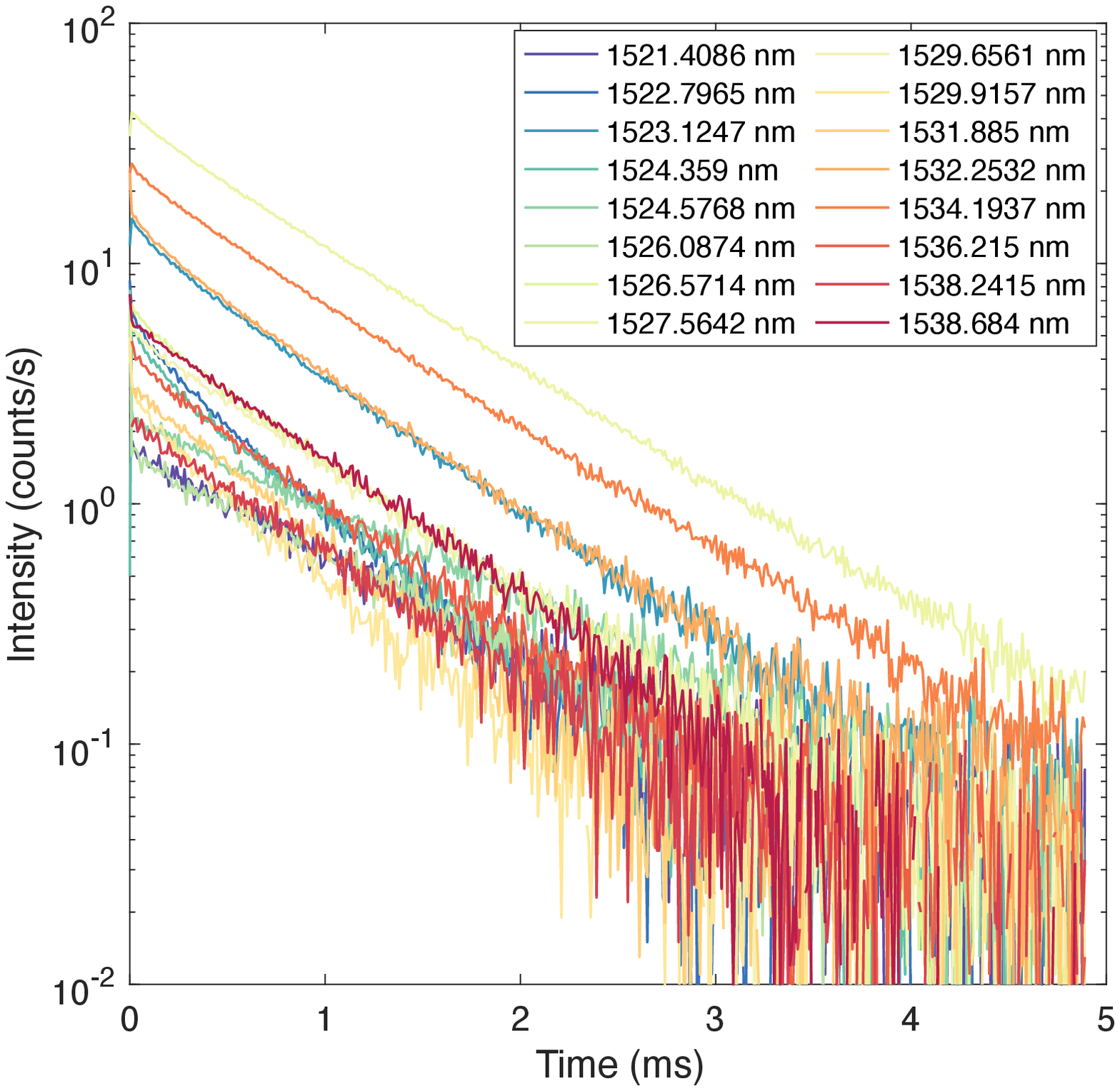}
    \caption{Photoluminescence decay for a multitude of resonances including biexponential decays.}
\label{fig:lifetimes}
\end{figure}

\newpage
Fig. \ref{fig:lifetimes} shows multiple time traces after subtracting their corresponding background.

\begin{figure}[h]
	\centering
    \includegraphics[width=1.0\columnwidth]{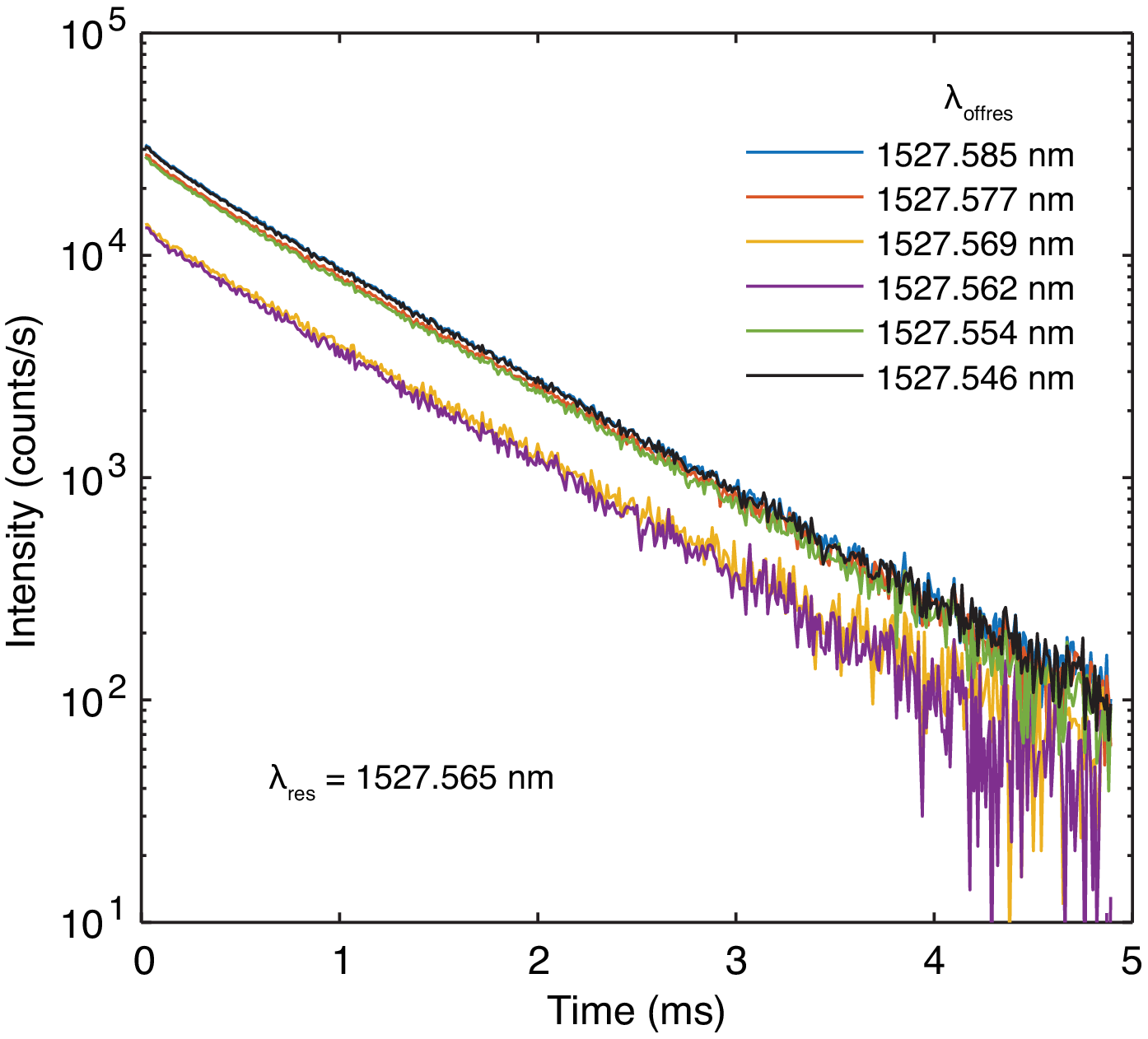}
    \caption{Photoluminescence decay with different backgrounds subtracted.}
\label{fig:diffoffres}
\end{figure}

To determine the dependence of the lifetime uncertainty on the chosen offresonant background decay, a measurement was carried out where six different background traces within ranging from f$_0-$\SI{2.5}{\giga\hertz} to f$_0+$\SI{2.5}{\giga\hertz} in steps of \SI{1}{\giga\hertz} have been subtracted from the decay measured in the centre of the inhomogeneous peak, represented by f$_0$. Figure \ref{fig:diffoffres} shows the resulting curves for \SI{1527.565}{\nano\metre}. 
Each trace is fitted with a single exponential and returns a lifetime and the standard error. The average of the six fitting errors is given in the column \textit{Average fitting error} and the standard deviation in the spread of the lifetimes in column \textit{Spread lifetimes} in table \ref{tab:lifetimeuncertainty}. 
This measurement was performed on a range of resonances which are presented in the column \textit{Resonant wavelength} in table \ref{tab:lifetimeuncertainty}. 
Following from these 8 resonances, the choice of background trace increases the standard error on average of by \num{2.13}.

%\vfill\null

\newpage
\renewcommand\arraystretch{0.5}
\begin{longtable}{
>{\RaggedRight}p{0.33\columnwidth} >{\RaggedRight}p{0.33\columnwidth} >{\RaggedRight}p{0.33\columnwidth}
}
    \caption{Lifetime uncertainties}\label{tab:lifetimeuncertainty}\\
        \toprule
         \thead{Resonant wavelength\\ (\si{\nano\metre})}& 
         \thead{Average fitting error\\ (\si{\micro\second})}&
         \thead{Spread lifetimes\\ (\si{\micro\second})}\\
         \midrule
         \num{1523.125} & \num{3.943}& \num{7.590}\\
         \num{1527.565} & \num{2.516}& \num{6.845}\\
         \num{1532.254} & \num{2.660}& \num{2.287}\\
         \num{1534.194} & \num{2.235}& \num{9.540}\\
         \num{1534.371} & \num{17.380}& \num{63.571}\\
         \num{1536.215} & \num{15.364}& \num{18.371}\\
         \num{1536.687} & \num{121.316}& \num{90.226}\\
         \num{1538.685} & \num{6.442}& \num{10.859}\\
        \bottomrule
\end{longtable}

\begin{figure}[h]
	\centering
	\includegraphics[width=1.0\columnwidth]{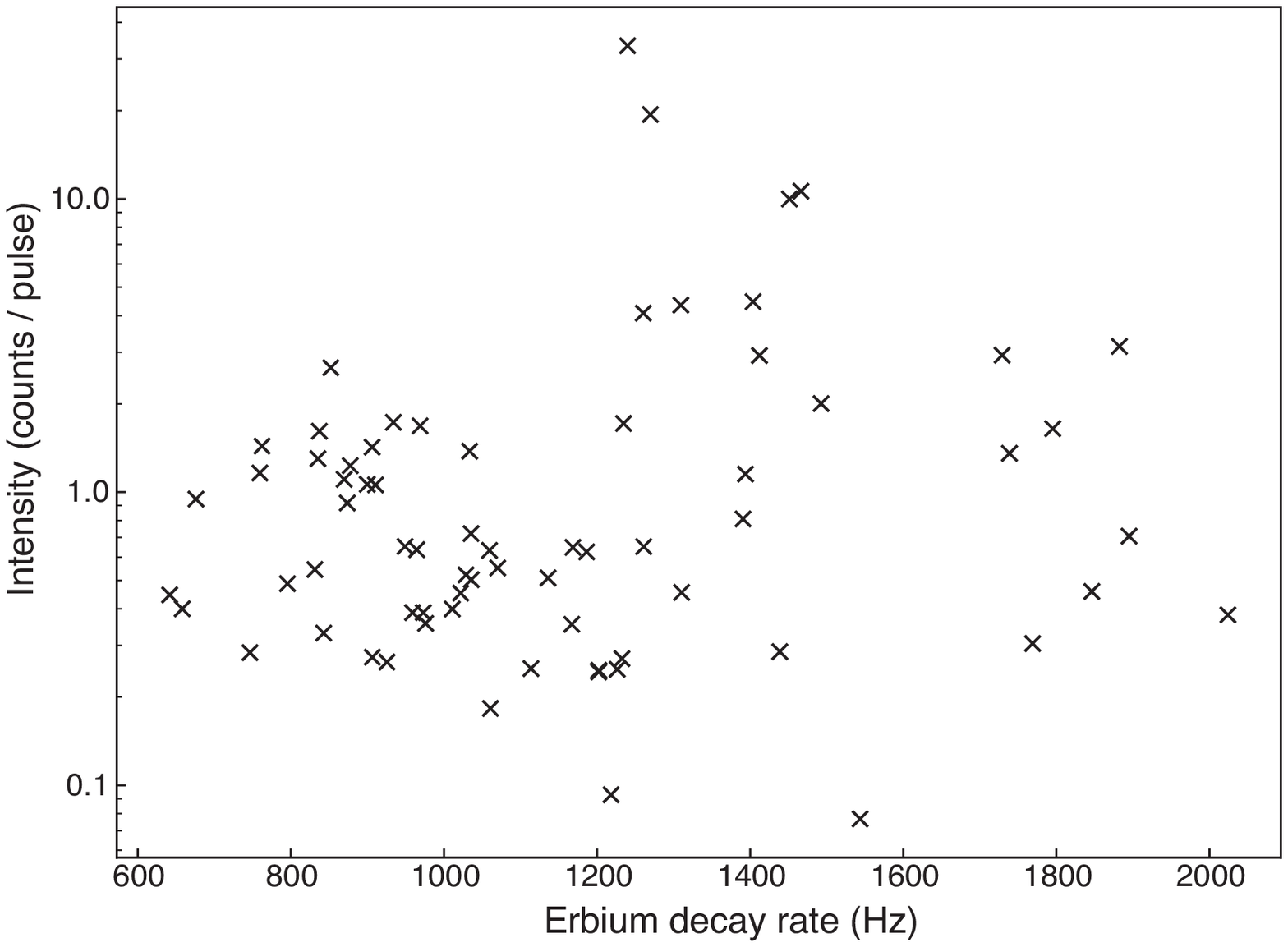}
	\caption{Intensity of each resonance as a function of the decay rate.}
	\label{fig:intensityvsdecay}
\end{figure}

In Fig. \ref{fig:intensityvsdecay} we compare the measured decay rate (1/lifetime) to the intensity of each line to analyze if the peak height is limited by the decay rate. It is apparent from the figure that no such correlation can be draw.

\newpage
\subsection{Instantaneous spectral diffusion}\label{sec:specdiffusion}

\begin{figure}[h]
	\centering
    \includegraphics[width=1.0\columnwidth]{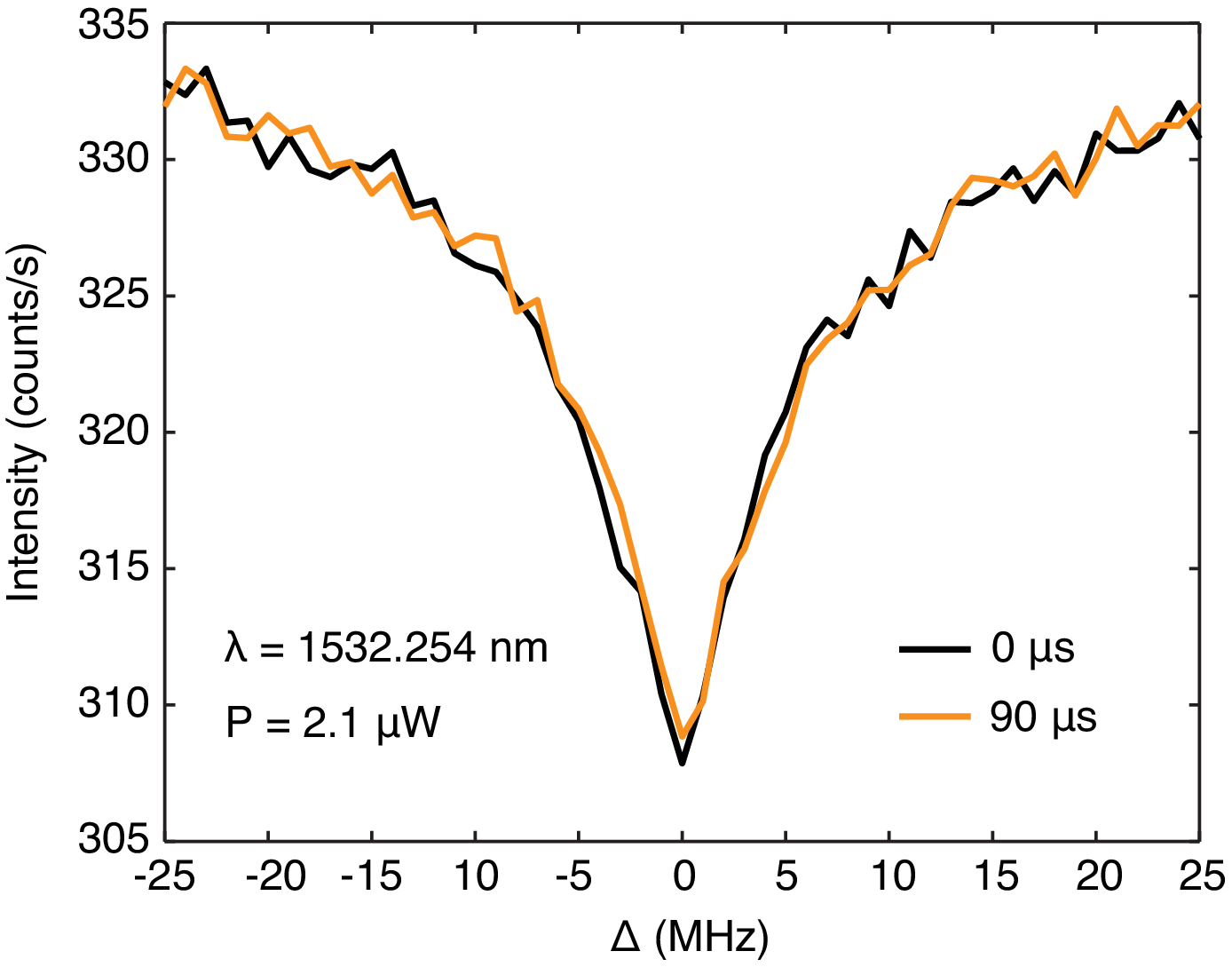}
    \caption{Overlapping plots of spectral hole burning with two different delays.}
    \label{fig:isd}
\end{figure}

To inspect if instantaneous spectral diffusion affects our homogeneous measurements, the spectral hole burning was performed with a \SI{90}{\micro\second} delay between the two pulses. The result can be seen in Fig. \ref{fig:isd} for a pump and probe pulse of \SI{10}{\micro\second} each.

%\clearpage
%\bibliographystyle{apsrev4-2}
%\bibliography{bibliography}

\end{document}